\documentclass[acmtog,authorversion,screen]{acmart}

\usepackage{bm}
\usepackage{tikz}
\usepackage{pifont}
\usepackage{enumitem}
\usepackage{multirow}
\usepackage{xcolor, colortbl}
\usepackage[norelsize, linesnumbered, ruled, lined, boxed, commentsnumbered]{algorithm2e}

\newcommand{\eg}{\textit{e.g.}}
\newcommand{\ie}{\textit{i.e.}}

\newcommand{\etc}{\textit{etc}}
\newcommand{\figref}[1]{Fig.~\ref{fig:#1}}
\newcommand{\tabref}[1]{Table~\ref{tab:#1}}
\newcommand{\secref}[1]{Sec.~\ref{sec:#1}}
\newcommand{\cmark}{\ding{51}}
\newcommand{\xmark}{\ding{55}} 

\definecolor{ourcolor}{HTML}{FFF2CC}

\copyrightyear{2025}
\acmYear{2025}
\setcopyright{cc}
\setcctype{by-nc}
\acmConference[SA Conference Papers '25]{SIGGRAPH Asia 2025 Conference
Papers}{December 15--18, 2025}{Hong Kong, Hong Kong}
\acmBooktitle{SIGGRAPH Asia 2025 Conference Papers (SA Conference Papers '25), December 15--18, 2025, Hong Kong, Hong Kong}
\acmDOI{10.1145/3757377.3763906}
\acmISBN{979-8-4007-2137-3/2025/12}

\newcommand{\PlainTitleText}{Bokeh Diffusion: Defocus Blur Control in Text-to-Image Diffusion Models}
\newcommand{\FancyTitle}{%
  \ifpdf
    \makebox[0pt][l]{\textcolor{white}{Bokeh}}%
    \includegraphics[height=12pt]{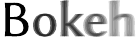}\hspace{-1pt}
  \else
    Bokeh
  \fi
  Diffusion: Defocus Blur Control in Text-to-Image Diffusion Models%
}

\makeatletter
\let\ACM@orig@mkbibcitation\@mkbibcitation
\def\@mkbibcitation{%
  \begingroup
    \renewcommand{\FancyTitle}{\PlainTitleText}%
    \ACM@orig@mkbibcitation
  \endgroup
}
\makeatother

\begin{document}

\title[\PlainTitleText]%
{\texorpdfstring{\FancyTitle}{\PlainTitleText}}

\author{Armando Fortes}
\affiliation{%
    \institution{S-Lab, Nanyang Technological University}
    \city{Singapore}
    \country{Singapore}
}
\email{armandol001@ntu.edu.sg}

\author{Tianyi Wei}
\affiliation{%
    \institution{S-Lab, Nanyang Technological University}
    \city{Singapore}
    \country{Singapore}
}
\email{tianyi.wei@ntu.edu.sg}

\author{Shangchen Zhou}
\affiliation{%
    \institution{S-Lab, Nanyang Technological University}
    \city{Singapore}
    \country{Singapore}
}
\email{sczhou@ntu.edu.sg}

\author{Xingang Pan}
\affiliation{%
    \institution{S-Lab, Nanyang Technological University}
    \city{Singapore}
    \country{Singapore}
}
\email{xingang.pan@ntu.edu.sg}

\begin{abstract}
Recent advances in large-scale text-to-image models have revolutionized creative fields by generating visually captivating outputs from textual prompts; however, while traditional photography offers precise control over camera settings to shape visual aesthetics---such as depth-of-field via aperture---current diffusion models typically rely on prompt engineering to mimic such effects. This approach often results in crude approximations and inadvertently alters the scene content. In this work, we propose Bokeh Diffusion, a scene-consistent bokeh control framework that explicitly conditions a diffusion model on a physical defocus blur parameter. To overcome the scarcity of paired real-world images captured under different camera settings, we introduce a hybrid training pipeline that aligns in-the-wild images with synthetic blur augmentations, providing diverse scenes and subjects as well as supervision to learn the separation of image content from lens blur. Central to our framework is our grounded self-attention mechanism, trained on image pairs with different bokeh levels of the same scene, which enables blur strength to be adjusted in both directions while preserving the underlying scene. Extensive experiments demonstrate that our approach enables flexible, lens-like blur control, supports downstream applications such as real image editing via inversion, and generalizes effectively across both Stable Diffusion and FLUX architectures.
\end{abstract}



\begin{teaserfigure}
    \centering
    \includegraphics[width=\textwidth]{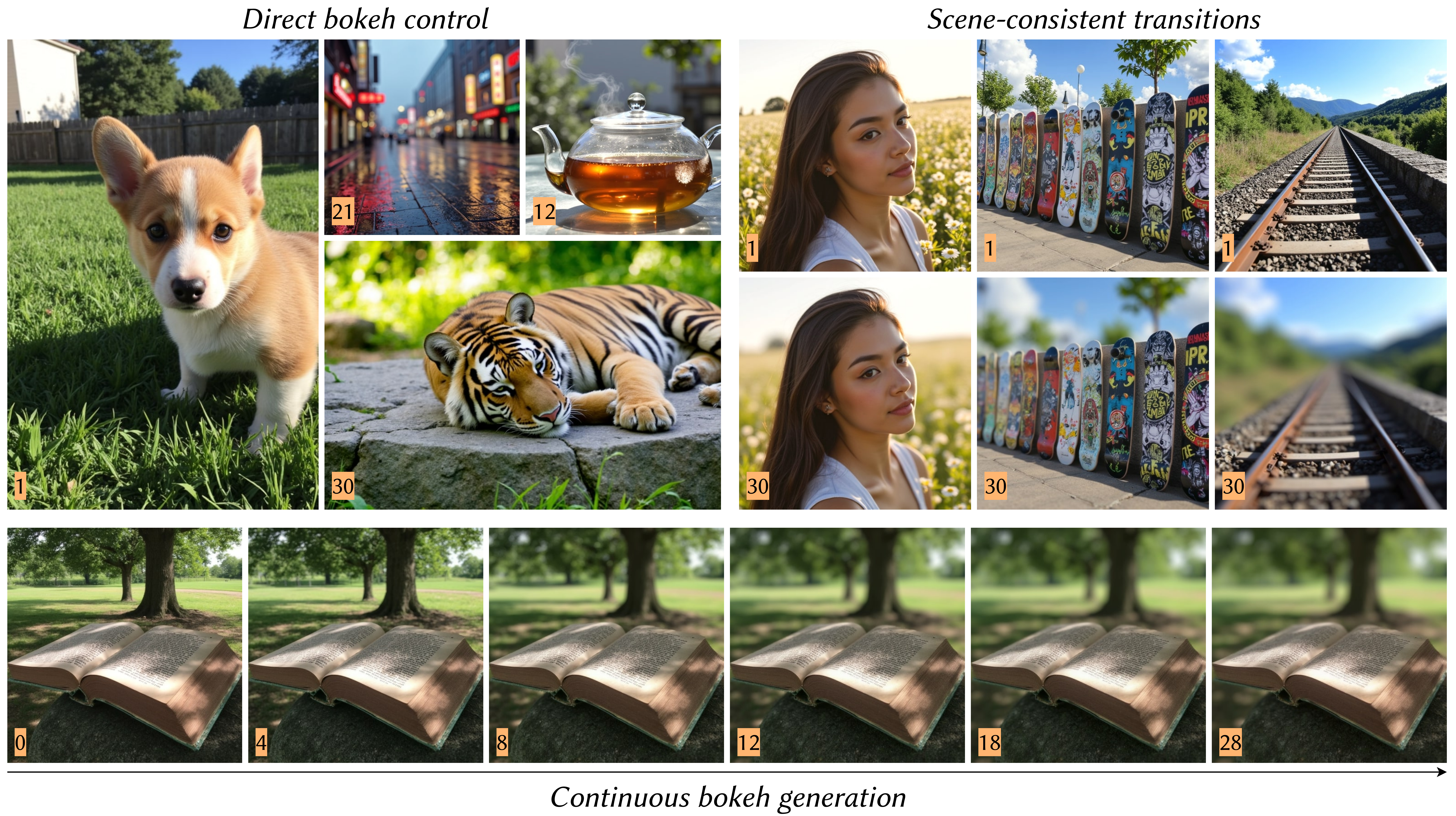}
    \caption{Bokeh Diffusion enables precise, scene-consistent bokeh control in text-to-image diffusion models. \emph{Top Left:} Images are conditioned on an explicit blur parameter to generate outputs ranging from sharp to strongly defocused (from 0 to 30). \emph{Top Right:} Generated scene content is consistent across different blur level conditions. \emph{Bottom:} Continuous control over blur strength produces smooth transitions across the bokeh range.}
    \Description{}
    \label{fig:teaser}
\end{teaserfigure}

\maketitle

\renewcommand{\shortauthors}{A. Fortes, T. Wei, S. Zhou, X. Pan}

\section{Introduction}
\label{sec:intro}

\begin{figure}[t]
    \centering
    \includegraphics[width=\columnwidth]{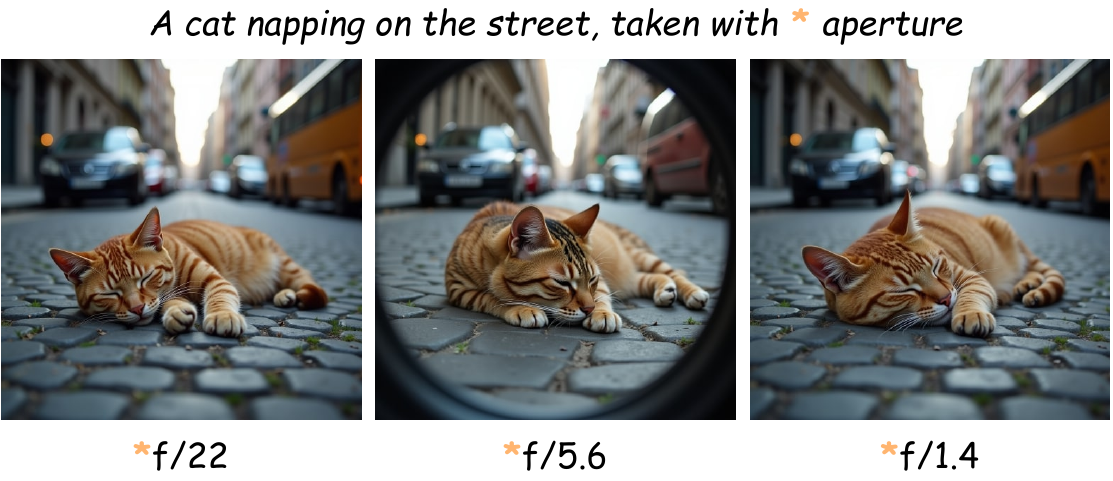}
    \caption{State-of-the-art text-to-image models~\cite{labsFLUX2024} do not understand textual photographic cues, and unintentionally alter image composition.}
    \Description{}
    \label{fig:flux-motivation}
\end{figure}

\begin{figure}[t]
    \centering
    \includegraphics[width=\columnwidth]{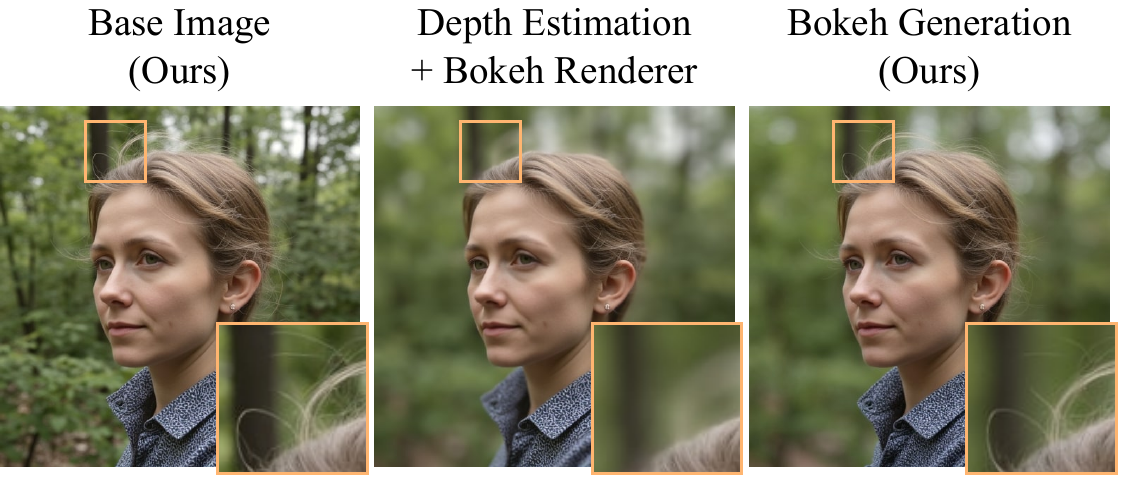}
    \caption{Comparison between our generative method and BokehMe~\cite{pengBokehMeWhenNeural2022}, which depends on external depth estimation models.}
    \Description{}
    \label{fig:comparison-ref}
\end{figure}

Photography has long offered a rich set of camera controls for shaping visual aesthetics. By varying the lens aperture and focal distance, photographers can decide whether a scene appears deeply in focus or softly blurred in the foreground or background~\cite{langford_basic_2015}. In professional workflows, these lens-based adjustments are integral to crafting the photograph's mood and composition~\cite{freeman2007eye}. Yet, in the rapidly evolving domain of text-to-image (T2I) generation, reproducing such camera effects is surprisingly difficult. Existing approaches typically rely on \emph{prompt engineering} (\eg, adding ``f/1.8 aperture'' to the text prompt), which often yields inaccurate approximations and unintentionally alters the scene content. As a result, changing apertures might yield not just a different bokeh but an entirely different object arrangement (see \figref{flux-motivation}).

Recent works have started to explore controlling camera intrinsics in image generation via camera tokens~\cite{fangCameraSettingsTokens2024}, but often still entangle depth-of-field with content changes. Others model camera settings along the temporal axis of a text-to-video framework to improve consistency~\cite{yuanGenerativePhotographySceneConsistent2024}, constantly requiring up to 7 frames per prompt at inference. This adds significant overhead and redundancy, particularly when only a single bokeh level is needed, and is not compatible with state-of-the-art T2I models. Moreover, despite the increased complexity, we find it still struggles to preserve fine scene details across large bokeh transitions.

In contrast, we propose a method for scene-consistent bokeh control directly within the T2I paradigm, offering greater efficiency and flexibility while fully leveraging the generative power of modern T2I backbones. Since real-world images with paired camera settings are scarce (\eg, identical viewpoint at f/1.8 and f/16), we align in-the-wild photographs (with EXIF metadata) and synthetic blur augmentations, allowing the model to learn from both authentic lens blur across diverse settings and structured focus--defocus pairs. Through this hybrid data pipeline, our model learns to better disentangle scene content from depth-of-field while closely mimicking lens-based defocus blur. Beyond dataset design, our framework uses lightweight cross-attention adapters to condition the diffusion process on a physical bokeh parameter, allowing fine-grained bokeh adjustments. Additionally, our specialized \emph{grounded self-attention} mechanism is specifically trained with image pairs, ensuring that blur changes respect the scene of a reference image. This design preserves content during bokeh manipulation and supports stable transitions between deep and shallow depth-of-field.

\begin{figure*}[t]
    \includegraphics[width=\linewidth]{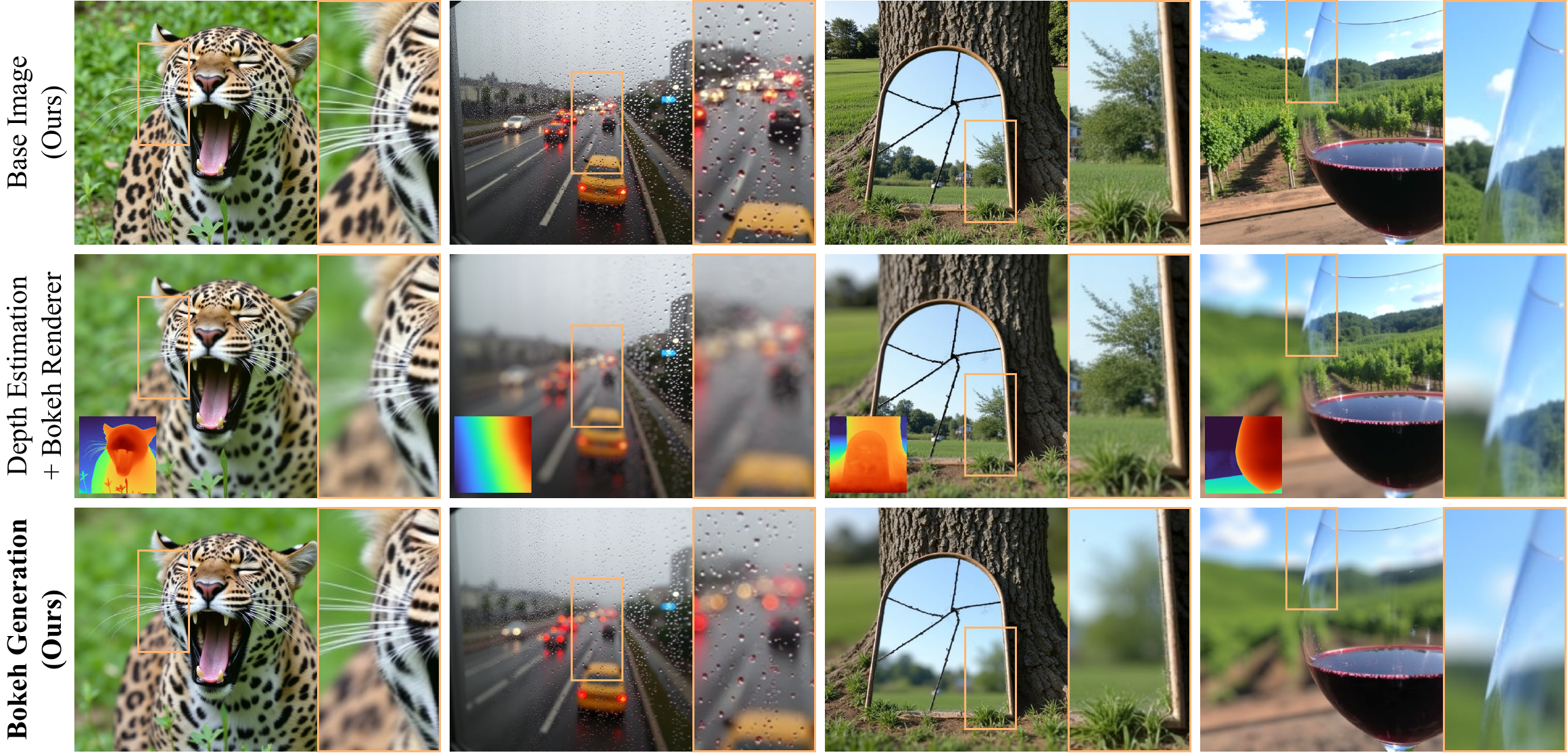}
    \caption{Additional comparison between our generative method and BokehMe~\cite{pengBokehMeWhenNeural2022}, which depends on external depth estimation models.}
    \Description{}
    \label{fig:bokehme}
\end{figure*}

Compared to post-processing approaches that add blur to images based on estimated depth maps~\cite{pengBokehMeWhenNeural2022}, our method offers several key advantages. First, it allows direct bokeh control during generation, letting users specify blur strength without post-processing. Second, given a generated image, our method supports adjusting the bokeh strength in both directions—either increasing or decreasing it—whereas post-processing methods struggle to remove blur. Third, by leveraging the generative prior of the pretrained diffusion model and our hybrid dataset, our method produces more natural bokeh effects, especially in challenging areas such as thin structures and translucent materials (see \figref{comparison-ref} and \figref{bokehme}).

In summary, (1) we propose Bokeh Diffusion, a \emph{scene-consistent bokeh control} framework for T2I diffusion, allowing physically motivated defocus blur changes without altering scene content; (2) we introduce a \emph{curated dataset} and \emph{hybrid training pipeline} leveraging EXIF-based cues from in-the-wild data and synthetic blur augmentations as contrastive examples; (3) we extensively evaluate our method and show that it consistently outperforms baselines across accuracy, consistency, and perceptual quality metrics; (4) we demonstrate the versatility of our approach by deploying it on both the UNet-based Stable Diffusion and the MMDiT-based FLUX architectures, achieving strong results across both; and (5) we explore additional applications such as real image editing via inversion.

\section{Related Work}
\label{sec:related}

\subsection{Text-to-image Diffusion}

Diffusion models~\cite{sohl-dicksteinDeepUnsupervisedLearning2015a, hoDenoisingDiffusionProbabilistic2020b} currently dominate text-to-image (T2I) generation due to their high sample quality and diversity. GLIDE~\cite{nicholGLIDEPhotorealisticImage2022a} introduced text-conditioned diffusion with a cascaded architecture for improved resolution, while Stable Diffusion~\cite{rombachHighResolutionImageSynthesis2022} shifted generation to a lower-dimensional latent space for efficient high-resolution synthesis. Other notable works include~\cite{sahariaPhotorealisticTexttoimageDiffusion2022a, baoOneTransformerFits2023a, podellSDXLImprovingLatent2023, chenPixArtaFastTraining2023, chenPixArtSWeaktoStrongTraining2024}. More recently, Stable Diffusion 3~\cite{esserScalingRectifiedFlow2024} and FLUX~\cite{labsFLUX2024} set the state-of-the-art, by scaling diffusion transformer~\cite{peeblesScalableDiffusionModels2023, baoAllAreWorth2023a} backbones with rectified flows~\cite{liuFlowStraightFast2022, lipmanFlowMatchingGenerative2022}. Despite their advancements in prompt fidelity and scene composition, these models struggle with fine-grained photographic controls, such as depth-of-field manipulation.

Recent works have also explored enhancing controllability. DreamBooth~\cite{ruizDreamBoothFineTuning2023a} enables subject-driven generation, while ControlNet~\cite{zhangAddingConditionalControl2023a} incorporates structural conditions like pose or edge maps. T2I-Adapter~\cite{mouT2IAdapterLearningAdapters2023} and IP-Adapter~\cite{yeIPAdapterTextCompatible2023a} facilitate reference image-driven generation. Further studies reveal that self- and cross-attention layers encode structural and semantic information, leveraged for editing~\cite{hertzPrompttoPromptImageEditing2022, caoMasaCtrlTuningFreeMutual2023a}, style adaptation~\cite{hertzStyleAlignedImage2024, wangInstantStyleFreeLunch2024}, and appearance transfer~\cite{alalufCrossImageAttentionZeroShot2024}.

\subsection{Camera-conditioned Diffusion Models}

Most existing works on camera control in diffusion models focus on extrinsic parameters, such as camera position and viewing angle~\cite{kumariCustomizingTexttoImageDiffusion2024, holleinViewDiff3DConsistentImage2024, chengLearningContinuous3D2024}, or trajectory in video generation~\cite{wangMotionCtrlUnifiedFlexible2023, heCameraCtrlEnablingCamera2024, sunDimensionXCreateAny2024, xiaoVideoDiffusionModels2024, xiaoTrajectoryAttentionFinegrained2025}. Regarding intrinsic camera control, \citet{voynovCurvedDiffusionGenerative2024} explore conditioning T2I diffusion on diverse optical geometries to simulate lens-induced distortions. Additionally, some methods model zoom implicitly~\cite{wangGenerativePowersTen2024, chengLearningContinuous3D2024}, while others perform prompt-based conditioning on camera settings~\cite{fangCameraSettingsTokens2024, nvidiaEdifyImageHighQuality2024}.

Camera Settings as Tokens (CSaT)~\cite{fangCameraSettingsTokens2024} introduces conditioning on intrinsic parameters such as aperture, but does not enforce scene consistency (\ie, changing aperture often alters the scene). While pairing CSaT with ControlNet~\cite{zhangAddingConditionalControl2023a} improves structural guidance, it does not guarantee consistent fine-grained scene content. Generative Photography~\cite{yuanGenerativePhotographySceneConsistent2024} addresses scene consistency by modeling intrinsic camera control along the temporal axis of a text-to-video (T2V) model. However, it relies on temporal modules and requires constantly generating several frames during inference, adding overhead and limiting single-image generation or editing. In contrast, our method achieves scene-consistent bokeh control entirely within the T2I paradigm. This allows direct integration into state-of-the-art T2I models with stronger generative capability, avoiding multi-frame inference overhead and redundancy, and supporting efficient, scalable deployment in both image generation and editing workflows.

\section{Method}
\label{sec:method}

\subsection{Background}
\label{sec:back}

\subsubsection{Diffusion models}
Diffusion models~\cite{sohl-dicksteinDeepUnsupervisedLearning2015a, hoDenoisingDiffusionProbabilistic2020b} learn to synthesize data samples from Gaussian noise. In the forward process, diffusion models gradually add noise to the data $x_0 \sim q\left(x_0\right)$, formalized by a Markov chain:
\begin{align}
    & q\left(x_{1:T}|x_0\right) = \prod_{t} q\left(x_t|x_{t-1}\right),\\
    & q\left(x_t|x_{t-1}\right) = \mathcal{N}\left(x_t ; \sqrt{\alpha_t} x_{t-1}, \beta_t \bm{I}\right),
\end{align}
where $\beta_t$ is the noise schedule and $\alpha_t = 1 - \beta_t$.

In the reverse process, data is synthesized by a Gaussian model that learns to approximate the denoising step:
\begin{equation}
    p_\theta\left(x_{t-1} | x_t\right) = \mathcal{N}\left(x_{t-1} ; \mu_\theta\left(x_t,t\right), \sigma_t^2\bm{I}\right).
\end{equation}
The mean $\mu_\theta\left(x_t,t\right)$ can be derived to learn a noise prediction network $\varepsilon_\theta$, which is often trained with MSE loss:
\begin{equation}
    \min_{\theta} \quad \mathbb{E}_{t, x_0, \varepsilon} \Vert \varepsilon - \varepsilon_\theta\left(x_t, t\right) \Vert_2^2,
\end{equation}
where $x_t = \sqrt{\overline{\alpha}_t} x_0 + \sqrt{1 - \overline{\alpha}_t}\varepsilon$ and $t \sim \{1, 2, \dots, T\}$. 

For conditional generation, we model the distribution $q\left(x_0 \vert c\right)$. The noise prediction network can be adapted by incorporating the condition $c$ during training, typically via the attention mechanism.

\subsubsection{Attention mechanism in T2I diffusion.}
Formally, attention can be expressed as a weighted sum of values $V$, with weights determined by the compatibility between a query $Q$ and keys $K$:
\begin{equation}
    \text{Attention}(Q, K, V) = \text{Softmax}\Bigl(\frac{QK^\top}{\sqrt{d}}\Bigr)V.
\end{equation}
In most T2I diffusion models~\cite{rombachHighResolutionImageSynthesis2022, sahariaPhotorealisticTexttoimageDiffusion2022a, podellSDXLImprovingLatent2023, chenPixArtaFastTraining2023, chenPixArtSWeaktoStrongTraining2024}, $Q$ is projected from the spatial features, while $K$ and $V$ are projected either from the same spatial features (self) or from text embeddings (cross).

\subsubsection{Thin lens model.}\label{sec:thinlens} As a camera captures a scene onto a 2D imaging plane, some parts may be in focus and others not. Given a circular aperture, the image of any point source not located at the focal distance is a small disk. This disk is often referred to as the circle of confusion (CoC), and its diameter $d$ depends on the scene and camera settings~\cite{lagendijkChapter14Basic2009, wangImplicitNeuralRepresentation2023b}. Specifically, $d$ can be calculated by:
\begin{equation}\label{eq:thin_lens}
    d = \frac{f^2}{\, N\left( S_1-f \right)\,} \frac{\,\lvert S_2-S_1 \rvert\,}{S_2},
\end{equation}
where $f$ is the focal length, $N$ is the aperture (\ie, F-number), $S_1$ is the focus distance, $S_2$ is the subject distance.

\subsubsection{Defocus blur rendering.}\label{sec:rendering} In blur rendering methods~\cite{pengInteractivePortraitBokeh2021, yangVirtualDSLRHigh2016}, the idea is to scatter each pixel in its neighboring areas where the distance between them is less than its blur radius. As discussed in~\cite{yangVirtualDSLRHigh2016, wadhwaSyntheticDepthoffieldSinglecamera2018, pengBokehMeWhenNeural2022}, given the disparity of a pixel, its blur radius is:
\begin{equation}\label{eq:radius}
    r=K\,\cdot\,\Delta disp,\quad \Delta disp = \left| \frac{1}{S_1}-\frac{1}{S_2} \right|\,.
\end{equation}
Such methods are capable of rendering realistic bokeh effects in depth-continuous areas but they tend to suffer from color-bleeding artifacts at depth discontinuities~\cite{pengBokehMeWhenNeural2022}. Recent rendering methods mainly focus on improving this limitation, \eg, using neural renderers to fix erroneous areas~\cite{pengBokehMeWhenNeural2022}, or layered occlusion-aware bokeh rendering~\cite{shengDrBokehDiffeRentiable2024}.

\subsection{Leveraging In-the-wild Image Augmentation}

In principle, real photographs with naturally occurring lens blur provide the most faithful representation of defocus blur in varied scenes. However, major limitations arise when relying exclusively on in-the-wild (ITW) data. First, images of the same scene captured at multiple camera settings (\eg, identical viewpoint at f/1.8 and f/16) are exceedingly rare. Consequently, the model can struggle to disentangle the defocus blur from the underlying scene arrangement, leading to local minima. Second, real images often exhibit shooting preference bias (\eg, portrait photographers opting for wide apertures, landscape photographers going for narrow ones). While there do exist datasets with multi-aperture captures of the same scene~\cite{abuolaim2020defocus, ignatov2020rendering, zhangSyntheticDefocusLookahead2019}, they are typically small in scale due to the manual effort required for such data collection. Moreover, these datasets tend to focus on static, inanimate scenes, since capturing precisely aligned shots with varying apertures is difficult when dynamic elements (\eg, people or animals) are present. This limited diversity makes them less suitable for training a generative model that must generalize across varied subjects, environments, and compositions.

To address these challenges, we build a hybrid training pipeline combining two complementary data sources. First, we curate a collection of in-the-wild photographs enriched with lens metadata, which offers diverse content and authentic photographic blur across many scenes (\secref{dataset-itw}). Second, we apply synthetic defocus blur augmentations to nearly all-in-focus images~\cite{yuanGenerativePhotographySceneConsistent2024} (\secref{dataset-synth}) for contrastive image pairs. This strategy enables the model to learn realistic defocus in diverse settings while also acquiring supervision to disentangle blur from the scene.

\begin{figure}[t]
    \centering
    \includegraphics[width=\linewidth]{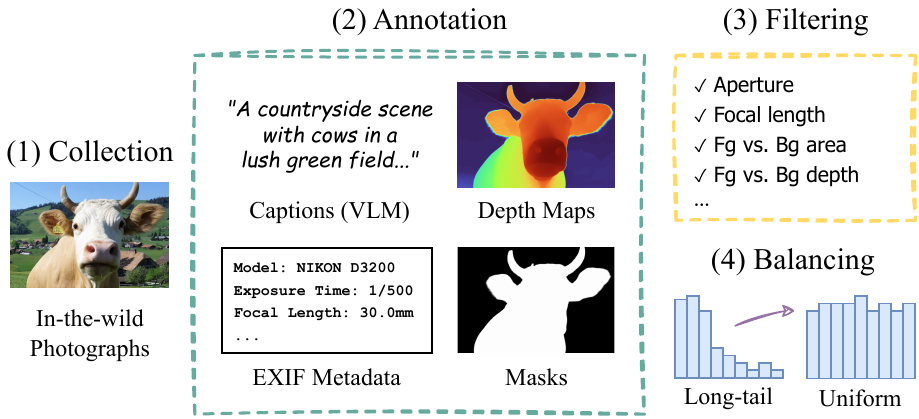}
    \vspace{-5mm}
    \caption{Data collection and curation pipeline. Cow \textcopyright{} Urs Rüegsegger (Flickr).}
    \Description{}
    \label{fig:dataset}
\end{figure}

\begin{figure}[t]
    \centering
    \includegraphics[width=\linewidth]{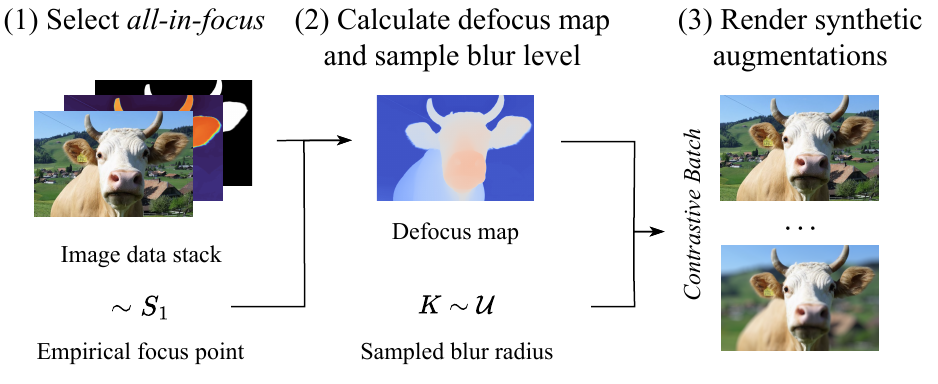}
    \vspace{-5mm}
    \caption{In-the-wild image augmentation. Cow \textcopyright{} Urs Rüegsegger (Flickr).}
    \Description{}
    \label{fig:aug}
\end{figure}

\subsubsection{In-the-wild images (\figref{dataset})}\label{sec:dataset-itw}

We curate a dataset of in-the-wild photographs from Flickr\footnote{\href{https://www.flickr.com/}{https://www.flickr.com/}}\footnote{\href{https://huggingface.co/datasets/madebyollin/megalith-10m}{madebyollin/megalith-10m}}. We select images licensed as public-domain with no copyright, and annotate each image with (a) estimated depth maps using \citet{bochkovskiiDepthProSharp2024}, (b) foreground masks produced via \citet{zhengBilateralReferenceHighResolution2024}, (c) EXIF metadata, and (d) text captions from multiple VLMs~\cite{chen2024sharegpt4v, chen2024internvl, xiao2024florence}. We filter out images lacking reliable lens info, unreasonable foreground area, small depth range, \etc. Finally, we perform balancing based on our physical defocus blur annotation obtained from the EXIF metadata. This step mitigates the strong long-tail bias in real photographs, providing a more uniform representation. The resulting dataset has around 15K samples, of which around 10\% are nearly all-in-focus and used for synthetic blur augmentation.

\subsubsection{Synthetic blur augmentation (\figref{aug})}\label{sec:dataset-synth}

We synthesize additional defocus examples by rendering artificial blur on nearly all-in-focus images using BokehMe~\cite{pengBokehMeWhenNeural2022}. Specifically, we estimate a per-pixel disparity map and then sample a blur parameter $K$ to generate contrastive ``focus--defocus'' states of the same image. These image pairs help the model avoid local minima by providing direct evidence that the same object layout can appear either sharp or defocused, and they are also essential for training our \emph{grounded self-attention} module, introduced in the following sections. Moreover, synthetic rendering allows us fine-grained control over the bokeh distribution as it allows us to continuously sample from the camera parameter space, unlike real datasets which are often constrained to a few discrete aperture settings.

\subsubsection{Mixed dataset training}

In practice, combining real ITW images with synthetic bokeh augmentations  provides an effective balance. To this end, it is necessary to ensure that bokeh conditioning is consistent across both data sources. As discussed in \secref{thinlens}, real-lens defocus can be estimated via the CoC diameter $d$ from the thin lens model, while synthetic blur rendering often uses a scalar $K$ as the pixel blur radius per unit disparity~\eqref{eq:radius}. Rewriting the CoC expression in terms of $\Delta disp$, the dependency on $S_2$ can be factored out. This yields a blur diameter proportional to disparity difference:
\begin{equation}
    d = \frac{f^2 S_1}{N(S_1 - f)} \cdot \Delta disp,
\end{equation}
which matches the linear disparity formulation used in synthetic rendering. Converting $d$ to a radius, the corresponding synthetic blur strength in the pixel space can be approximated by:
\begin{equation}
    K 
  \;\approx\;
  \frac{f^2 S_1}{\,2 N\,(S_1 - f)\,} \cdot \text{pixel\_ratio},
\end{equation}
where $f$ and $N$ are obtained from EXIF metadata, $S_1$ is calculated based on heuristics, and pixel\_ratio accounts for different camera sensor sizes and image resolution. Empirically, we find that setting $S_1$ to the median depth of the nearer 50\% of foreground pixels yields an acceptable approximation for most images. This formulation allows us to compute a physically grounded blur parameter $K$ for real images, aligned with the synthetic rendering domain. As a result, we can train the model using a unified defocus representation.

\begin{figure*}[t!]
    \centering
    \includegraphics[width=\linewidth]{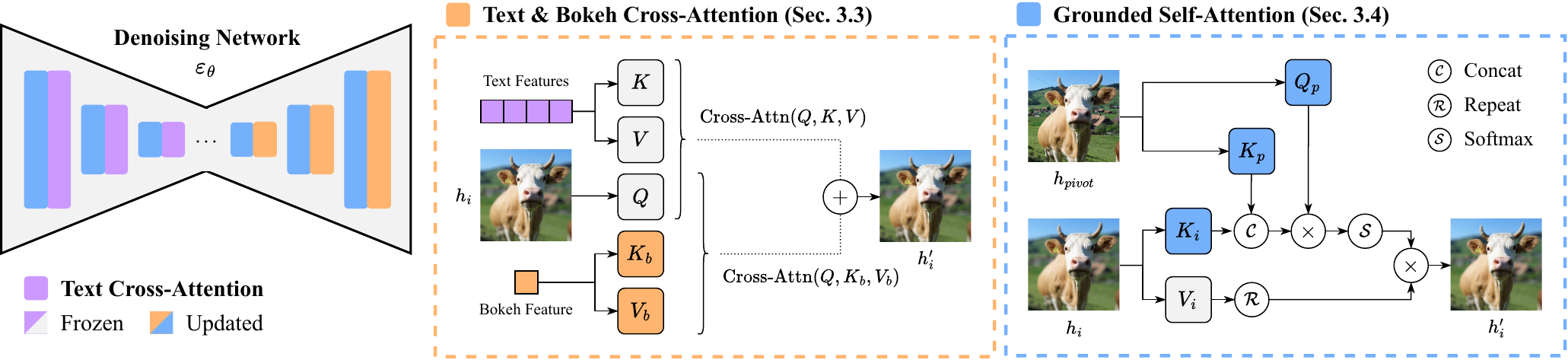}
    \caption{Overview of the conditioning and scene-consistency mechanisms. \textit{Bokeh Cross-Attention} (\secref{bokeh}), applied at deeper layers of the diffusion model, injects a defocus conditioning signal. \textit{Grounded Self-Attention} (\secref{scene}) ensures fine-grained scene consistency across different defocus levels, where the pivot image's queries anchor the scene structure, while shared keys allow defocus blur disentanglement from scene composition. Cow \textcopyright{} Urs Rüegsegger (Flickr).}
    \Description{}
    \label{fig:method}
\end{figure*}

\subsection{Defocus Blur Conditioning}\label{sec:bokeh}

Inspired by previous work~\cite{yeIPAdapterTextCompatible2023a}, we employ decoupled cross-attention to inject our defocus blur condition into the existing T2I diffusion pipeline. Specifically, for each cross-attention layer in the decoder of the UNet, we add a new pair of key-value projections $W'_k, W'_v$ tied to a lightweight bokeh feature $c_b$. This feature is obtained by passing the scalar defocus parameter through a small MLP. The original text-based keys and values ($K, V$) remain frozen. Meanwhile, the new bokeh cross-attention is computed with the same query $Q$ but keys and values $K_b=c_b W'_k,\ V_b=c_b W'_v$, then linearly combined with the text cross-attention output:
\begin{equation}
h^{\text{new}}=\text{Attention}(Q,K,V)\,+\,\lambda\,\cdot\,\text{Attention}(Q,K_b,V_b).
\end{equation}
Empirically, attaching our adapter to the deeper cross-attention layers sufficiently shapes the final texture while minimizing leakage. \figref{method} illustrates how text and bokeh features are combined for defocus control without disrupting scene semantics. In addition, we adapt the attention localization strategy from \citet{xiaoFastComposerTuningFreeMultiSubject2023} to steer bokeh attention to the background region during training.

\subsection{Imposing Scene-consistent Bokeh Generation}\label{sec:scene}

\subsubsection{Grounded self-attention.}
While bokeh cross-attention effectively controls defocus blur for a single image, we observe unwanted content shifts when applying different blur levels to the same scene (\ie, same text prompt and random seed). To address this, we introduce our \emph{grounded self-attention} mechanism (see \figref{method}). This design follows prior analyses of attention in image diffusion models showing that queries primarily encode spatial information, keys act as the linking mechanism to features, and values capture appearance and style~\cite{alalufCrossImageAttentionZeroShot2024, tewel2025addit}. Leveraging these roles, we designate a random \emph{pivot} image in each contrastive batch as a structural anchor for the other images during training.

Formally, let $K_{\text{tgt}},\,V_{\text{tgt}}$ be the key and value projections from the deep features $h_{\text{tgt}}$ of target image $I_{\text{tgt}}$, and let $Q_{\text{piv}},\,K_{\text{piv}}$ be the query and key projections corresponding to a \emph{pivot} image $I_{\text{piv}}$. We replace the usual self-attention operation for $h_{\text{tgt}}$ with:
\begin{equation}
    \text{Attention}\bigl(Q_{\text{piv}},\,[K_{\text{tgt}},K_{\text{piv}}],\,[V_{\text{tgt}},V_{\text{tgt}}]\bigr),
\end{equation}
where $[\,\cdot,\cdot\,]$ indicates concatenation. In other words, the pivot’s query $Q_{\text{piv}}$ \emph{anchors} the structural layout, while the concatenated keys $[K_{\text{tgt}},K_{\text{piv}}]$ allow the attention mechanism to reason over the pivot’s intended structure and the target’s current layout, enabling more flexible correspondence. The values are taken only from $V_{\text{tgt}}$ (duplicated for dimensionality), so the resulting features are always drawn from the target, inheriting its appearance and defocus.

To enable this cross-image guidance, we match the noise and timesteps for $I_{\text{tgt}}$ and $I_{\text{piv}}$ during training, so both images undergo synchronized diffusion. Consequently, the backward (denoising) process for $I_{\text{tgt}}$ can be expressed as:
\begin{equation}
    z_{t-1} \;\leftarrow\; \varepsilon_{\theta} \bigl(z_t,\,c,\,c_b,\,t;\,\{Q^t_{\text{piv}},K^t_{\text{piv}}\}\bigr),
\end{equation}
where $z_{t-1}$ denotes the predicted latent at timestep $(t-1)$, $c$ is the textual context, $c_b$ the bokeh condition, and $\{Q^t_{\text{piv}},K^t_{\text{piv}}\}$ are the pivot’s queries and keys at timestep $t$.

\subsubsection{Color transfer.} Although our model mitigates most inconsistencies, we observe occasional biases in lighting or color between different defocus levels. To address this, we apply a post-processing step that aligns the color distribution of the target bokeh image with the pivot, inspired by the classic color-transfer method of~\citet{reinhardColorTransferImages2001}.
We operate in the $L^*a^*b^*$ color space and update each channel $C$ of image \(I_{\text{tgt}}\) as:
\begin{equation}
    I_{\text{tgt}}
    \;\leftarrow\;
    \frac{\sigma_{\text{piv}}}{\sigma_{\text{tgt}}} 
    \,\bigl(I_{\text{tgt}} - \mu_{\text{tgt}}\bigr)
    \;+\;\mu_{\text{piv}},
\end{equation}
thus matching channel-wise first- and second-order statistics. This preserves perceptual sharpness relationships yet harmonizes lighting and color across the image pair.

\setlength{\tabcolsep}{5pt}
\begin{table*}[t]
\caption{Quantitative comparison against pretrained and finetuned baselines. \emph{Accuracy} is measured by the Laplacian variance trend correlation to the reference (LVCorr) and GPT-4o estimations. \emph{Consistency} is measured by LPIPS, DINOv2, and DreamSim (where closer to the reference is better), as well as GPT-4o scores. \emph{Quality} is evaluated using ImageReward. The best result is showcased in \textbf{bold}, and the second-best \underline{underlined}.}
\centering
\begin{tabular}{llccccccc}
    \toprule
    \multirow{2}{*}{Type} & \multirow{2}{*}{Method} & \multicolumn{2}{c}{Accuracy} & \multicolumn{4}{c}{Consistency} & Quality \\ \cmidrule(lr){3-4} \cmidrule(lr){5-8} \cmidrule(lr){9-9}
     &  & LVCorr$\,\uparrow$ & GPT-4o$\,\uparrow$ & LPIPS & DINOv2 & DreamSim & GPT-4o$\,\uparrow$ & ImageReward$\,\uparrow$ \\ \midrule
    \textit{Reference} & BokehMe~\cite{pengBokehMeWhenNeural2022} & 1.000 & 0.941 & 0.075 & 0.994 & 0.012 & 96.65 & -- \\ \midrule
    \multirow{2}{*}{Pretrained} & SD1.5~\cite{rombachHighResolutionImageSynthesis2022} & -0.206 & 0.362 & 0.314 & 0.966 & 0.099 & 68.75 & 1.065 \\
     & FLUX~\cite{labsFLUX2024} & -0.033 & 0.336 & 0.177 & 0.983 & 0.058 & 78.25 & \textbf{1.280} \\ \midrule
    \multirow{5}{*}{Finetuned} & CSaT~\cite{fangCameraSettingsTokens2024} & -0.044 & 0.628 & 0.155 & 0.961 & 0.082 & 73.00 & 0.509 \\
     & CSaT + ControlNet~\cite{fangCameraSettingsTokens2024} & 0.498 & 0.784 & 0.178 & 0.972 & 0.071 & 67.25 & 0.335 \\
     & GenPhoto~\cite{yuanGenerativePhotographySceneConsistent2024} & 0.796 & 0.826 & 0.197 & 0.949 & 0.066 & 86.50 & -1.471 \\
    \cmidrule(lr){2-9} \rowcolor{ourcolor} \cellcolor{white}{} & \textbf{Bokeh Diffusion (SD1.5)} & \underline{0.904} & \underline{0.960} & \textbf{0.080} & \textbf{0.992} & \textbf{0.017} & \underline{94.75} & 0.988 \\
     \rowcolor{ourcolor} \cellcolor{white}{} & \textbf{Bokeh Diffusion (FLUX)} & \textbf{0.917} & \textbf{0.985} & \underline{0.052} & \textbf{0.996} & \underline{0.005} & \textbf{98.16} & \underline{1.093} \\
    \bottomrule
\end{tabular}
\label{tab:evaluation}
\end{table*}

\setlength{\tabcolsep}{1.8pt}
\begin{table}[t]
\caption{Additional evaluations of our method (with FLUX backbone). Results are reported under continuous bokeh control with randomly sampled blur levels, broken down by domain (objects, animals, and people).}
\centering
\begin{tabular}{lcccccc}
    \toprule
    \multirow{2}{*}{Subject} & \multicolumn{2}{c}{Accuracy} & \multicolumn{4}{c}{Consistency} \\ 
    \cmidrule(lr){2-3} \cmidrule(lr){4-7}
    & LVCorr & GPT-4o & LPIPS & DINOv2 & DreamSim & GPT-4o \\ \midrule
    Person ($\times$5)  & 0.921 & 0.928 & 0.030 & 0.994 & 0.012 & 95.00 \\
    Object ($\times$7) & 0.909 & 0.919 & 0.065 & 0.992 & 0.012 & 94.94 \\
    Animal ($\times$8) & 0.899 & 0.921 & 0.047 & 0.995 & 0.005 & 97.25 \\ \midrule
    Global avg. & 0.908 & 0.922 & 0.049 & 0.994 & 0.009 & 95.88 \\
    \bottomrule
\end{tabular}
\label{tab:continuous}
\end{table}

\subsection{Bokeh Diffusion in MMDiT Architecture}

Recent state-of-the-art T2I models (\eg, FLUX~\cite{labsFLUX2024}) replace the traditional UNet-based denoising network with a \emph{Multimodal Diffusion Transformer} (MMDiT)~\cite{esserScalingRectifiedFlow2024}. In FLUX, the MMDiT backbone performs joint self-attention over concatenated text and image tokens within a single, unified operation:
\begin{equation}
\begin{aligned}
\text{Attention}\bigl([Q_{\text{txt}},\operatorname{R}(Q_{\text{img}})],\,
 [K_{\text{txt}},\operatorname{R}(K_{\text{img}})],\,[V_{\text{txt}},V_{\text{img}}]\bigr)
\label{eq:flux_sa}
\end{aligned}
\end{equation}
where we use $\operatorname{R}(\cdot)$ to denote application of Rotary Positional Embedding (RoPE)~\cite{su2024roformer}. RoPE is injected only into image queries and keys, while text tokens keep all-zero positional encoding.

Since MMDiT no longer separates self- from cross-attention, we insert \emph{both} bokeh conditioning and pivot-based grounding directly into its joint attention module.  
Let $Q_{\text{piv}}=[\,Q^{\text{piv}}_{\text{txt}},\operatorname{RoPE}(Q^{\text{piv}}_{\text{img}})\,]$ denote the concatenated pivot queries, and define $K_{\text{tgt}},K_{\text{piv}},V_{\text{tgt}}$ analogously, the Bokeh Diffusion attention modules from \secref{bokeh} and \secref{scene} are jointly defined in FLUX as:
\begin{equation}
\begin{aligned}
\text{Attention}\bigl(Q_{\text{piv}},\,[K_{\text{tgt}},K_{\text{piv}}],\,[V_{\text{tgt}},V_{\text{tgt}}]\bigr)\\
{}+\lambda\,\text{Attention}\bigl(Q_{\text{piv}},K_b,V_b\bigr).
\end{aligned}
\end{equation}
This adaptation makes our framework fully compatible with FLUX.
\section{Experiments}
\label{sec:experiments}

\subsection{Scene-consistent Defocus Blur Control}

\subsubsection{Comparison with baselines.}

We compare our method against two pretrained text-to-image (T2I) diffusion models, SD1.5~\cite{rombachHighResolutionImageSynthesis2022} and FLUX~\cite{labsFLUX2024}, as well as two related methods that finetune diffusion models for photographic controls, Camera Settings as Tokens (CSaT)~\cite{fangCameraSettingsTokens2024} and Generative Photography (GenPhoto)~\cite{yuanGenerativePhotographySceneConsistent2024}. As a reference, we also include blur renderings from BokehMe~\cite{pengBokehMeWhenNeural2022}.

\paragraph{Quantitative evaluation.} We benchmark on 20 diverse prompts, each with 5 different bokeh conditions, totaling 100 samples.
The quantitative evaluation focuses on three aspects:
\begin{itemize}
    \item \emph{Accuracy}: We measure how each method’s bokeh aligns with the target blur parameter via CorrCoef~\cite{yuanGenerativePhotographySceneConsistent2024}. Specifically, we compute the Laplacian variance trend for images in a scene under different blur conditions and obtain the Pearson correlation with the reference trend (LVCorr). A higher correlation indicates more faithful defocus control. Additionally, we employ GPT-4o~\cite{openaiGPT4oSystemCard2024} to estimate the bokeh level of a batch of images and calculate the Pearson correlation with its conditioning sequence.
    \item \emph{Consistency}: We examine whether the blur level causes unintended shifts in the scene. We use LPIPS~\cite{zhangUnreasonableEffectivenessDeep2018}, DINOv2~\cite{oquab2024dinov}, and DreamSim~\cite{fu2023dreamsim} to quantify how distant images with same prompt but different blur conditions are. It is important to note that lower is not always better, since the change in depth-of-field should affect perceptual distance to some extent. We consider best the closest result to the reference. Additionally, we employ GPT-4o~\cite{openaiGPT4oSystemCard2024} to rate how consistent a batch of images remain despite depth-of-field variations.
    \item \emph{Quality}: We use ImageReward~\cite{xuImageRewardLearningEvaluating2023a}, a general-purpose T2I preference reward model, to assess the perceptual quality and prompt alignment of generated images.
\end{itemize}

\tabref{evaluation} shows that our approach outperforms all baselines by a significant margin. Pretrained T2I models show low accuracy, reflecting poor depth-of-field control. Furthermore, they also suffer scene shifts, as evidenced by the consistency scores. While recent finetuned methods improve upon the pretrained backbones in those two dimensions, they still fall short of our results. In terms of generation quality, our method achieves scores comparable to its respective pretrained backbone, indicating that our framework introduces scene-consistent bokeh control while largely maintaining the original generation quality. Compared to GenPhoto, which operates within a T2V framework and relies on batched generation with several frames, our method is built entirely within the T2I paradigm, allowing our model to be integrated into state-of-the-art T2I systems, inheriting their generation quality.

Additionally, we evaluate our method with randomly sampled blur levels in Table~\ref{tab:continuous}, showing that performance remains consistent with the discrete-level evaluation used for comparison with the baselines. We further report per-domain results, which confirm robustness across different content types.

\begin{figure*}[p]
    \includegraphics[width=0.88\linewidth]{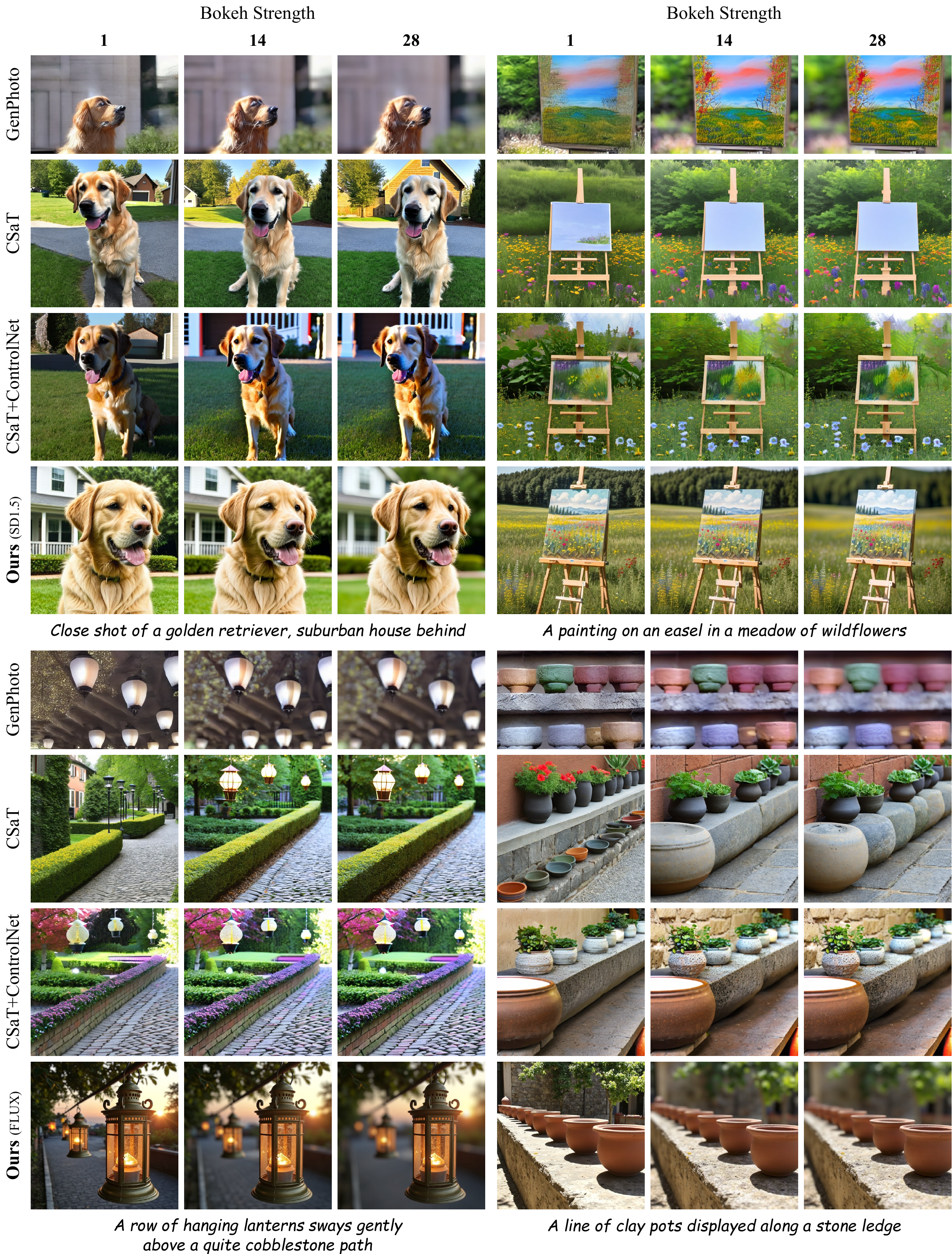}
    \caption{Visual comparison with related generative models conditioned on photographic controls CSaT~\cite{fangCameraSettingsTokens2024} and GenPhoto~\cite{yuanGenerativePhotographySceneConsistent2024}.}
    \Description{}
    \label{fig:comparison}
\end{figure*}

\begin{figure*}[t]
    \centering
    \begin{minipage}{0.48\linewidth}
        \centering
        \includegraphics[width=\linewidth]{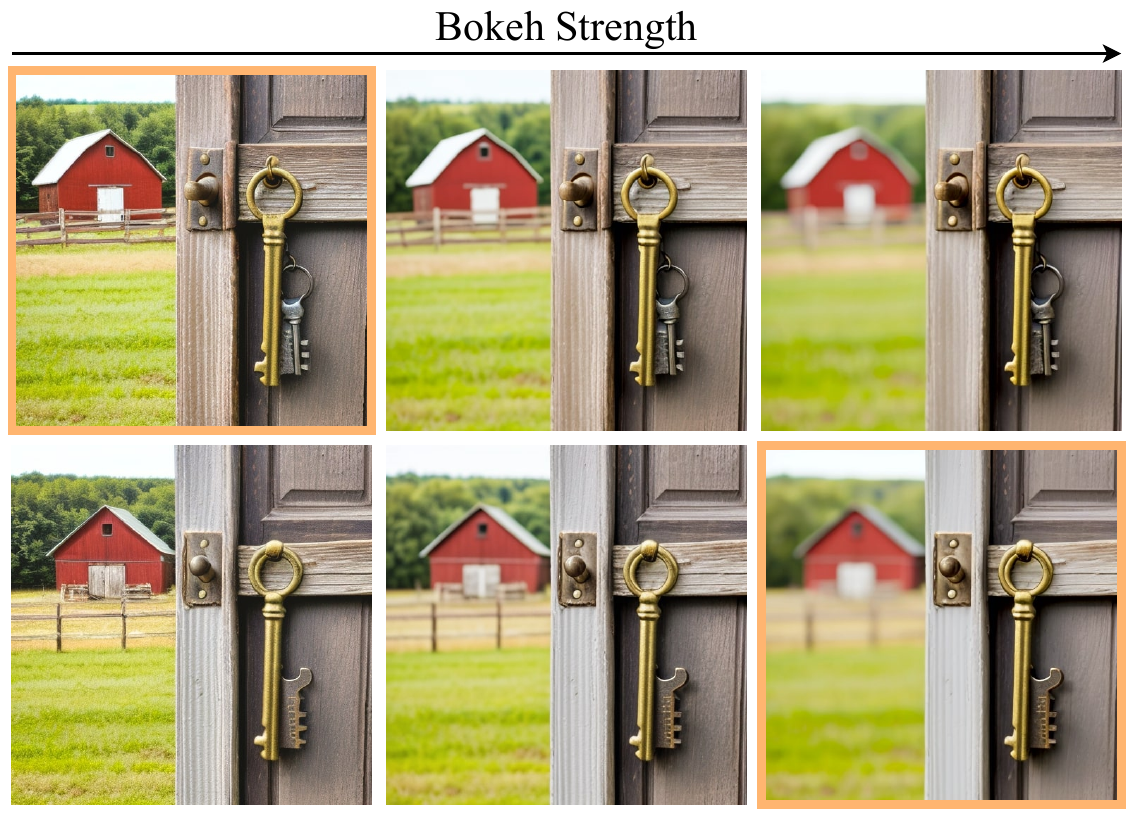}
        \caption{Generated image editing with Bokeh Diffusion (SD1.5). By selecting a pivot image (highlighted in each row) with a given bokeh level, the model can increase or decrease defocus while maintaining scene composition.}
        \label{fig:editing}
    \end{minipage}
    \hfill
    \begin{minipage}{0.48\linewidth}
        \centering
        \includegraphics[width=\linewidth]{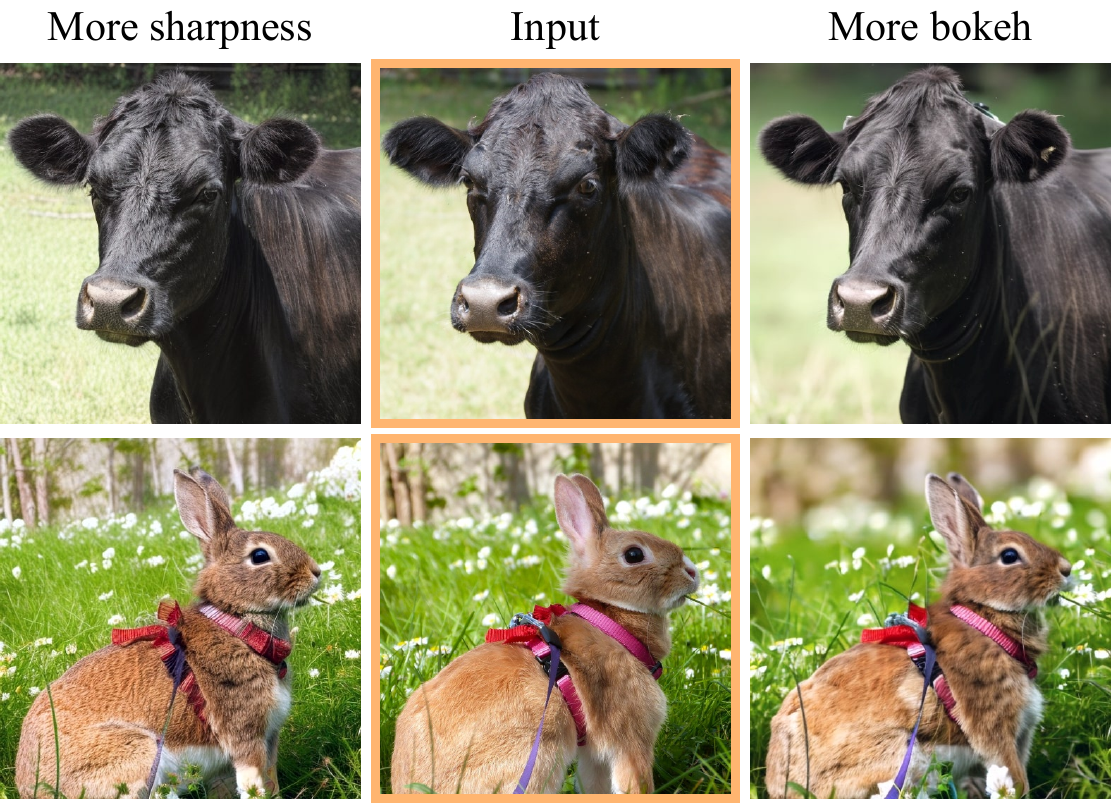}
        \caption{Real image editing with Bokeh Diffusion (SD1.5). A real photograph (highlighted) is inverted and denoised with a new bokeh condition. Inputs: top row \textcopyright{} Stephen Rahn (Flickr), bottom row \textcopyright{} Lo (Flickr).}
        \Description{}
        \label{fig:inversion}
    \end{minipage}
\end{figure*}

\paragraph{Qualitative evaluation.} As shown in \figref{comparison}, CSaT often alters the scene layout when changing bokeh levels, due to entangled conditioning. Even when combined with ControlNet~\cite{zhangAddingConditionalControl2023a}, as proposed by the authors, consistency is only coarse and the underlying content can still vary significantly. This inconsistency also affects the intended defocus effect, as the perception of bokeh depends on consistent subject-background relationships. GenPhoto improves scene stability by operating in the T2V domain with batched inference. However, low-bokeh results often appear overly textured or flat (\eg, lantern and painting scenes). It also exhibits subtle inconsistencies in content preservation, particularly in fine details and color distribution (\eg, clay pot and painting cases). In contrast, our method generates the full bokeh range accurately with high perceptual quality, while preserving scene consistency.

\paragraph{User study.} \tabref{user} summarizes the results of a user study conducted with 20 participants. Each participant viewed 15 scenes, where outputs from four anonymized methods were shown as image sequences in a grid (progressing from sharp to blurred, left to right), equivalent to anonymized versions of the grids in \figref{comparison}. For each scene, participants selected their preferred method along three criteria: (i) Bokeh Control (accurate sharp-to-defocused progression), (ii) Scene Consistency (stable content across blur levels), and (iii) Text-Image Alignment (faithfulness to the prompt). An example interface is provided in the supplementary material. Across all dimensions, users showed a clear preference for our method.

\setlength{\tabcolsep}{5pt}
\begin{table}[t]
\caption{User preference rates.}
\centering
\begin{tabular}{l|cccc}
    \toprule
    & CSaT & CSaT+CN & GenPhoto & \cellcolor{ourcolor}{Ours} \\ \midrule
    Bokeh Control & \underline{4.00\%} & 2.67\% & 2.33\% & \cellcolor{ourcolor}{\textbf{91.00\%}} \\
    Scene Consistency & 2.33\% & 2.33\% & \underline{2.67\%} & \cellcolor{ourcolor}{\textbf{92.67\%}} \\
    Text-Image Align. & \underline{13.00\%} & 10.00\% & 3.00\% & \cellcolor{ourcolor}{\textbf{74.00\%}} \\
    \bottomrule
\end{tabular}
\label{tab:user}
\end{table}

\subsubsection{Image editing.}
Our framework accommodates two usage scenarios: 
\emph{(i)} Unbounded bokeh generation, where a user can directly sample an image at the desired blur level using the conventional self-attention mechanism. \emph{(ii)} Grounded bokeh generation, where a pivot image is chosen to anchor the scene whose bokeh the user wants to adjust. Since we randomly pick a pivot in each contrastive batch during training, the grounded self-attention module in our model learns to either add or remove blur relative to that pivot. Consequently, at inference time we can choose any pivot image with a given bokeh level and request a new target level. \figref{editing} demonstrates how using either a sharper or blurrier pivot yields results matching the requested bokeh parameter without changing its contents. Additionally, we also support editing real images by first inverting them via a diffusion inversion procedure~\cite{songDenoisingDiffusionImplicit2020}, then simultaneously denoising two latents: one to reconstruct the original image (used as pivot) and one to apply a new bokeh setting. \figref{inversion} shows how our approach can natively add or reduce blur in a real image with minimal artifacts through inversion.

\subsection{Ablation Study}

We investigate two main aspects of our pipeline: training data composition and the method for imposing consistency.

\setlength{\tabcolsep}{1.8pt}
\begin{table}[t]
\caption{Quantitative ablation study on SD1.5 version.}
\centering
\begin{tabular}{lcccccc}
    \toprule
    \multirow{2}{*}{Method} & \multicolumn{2}{c}{Dataset} & \multicolumn{2}{c}{Accuracy} & \multicolumn{2}{c}{Consistency} \\ \cmidrule(lr){2-3} \cmidrule(lr){4-5} \cmidrule(lr){6-7}
    & ITW & SYN & LVCorr & GPT-4o & LPIPS & GPT-4o \\ \midrule
    \textit{Reference} & -- & -- & 1.000 & 0.941 & 0.075 & 96.65 \\ \midrule
    Pretrained SD1.5 & -- & -- & -0.206 & 0.362 & 0.314 & 68.75 \\ \midrule
    \multirow{3}{*}{+ Bokeh Cross-Attn} & \cmark & \xmark & 0.705 & 0.304 & 0.098 & 90.50 \\
     & \xmark & \cmark & 0.698 & 0.618 & 0.202 & 80.00 \\
     & \cmark & \cmark & 0.722 & 0.690 & 0.171 & 80.00 \\ \midrule
    + Grounded Self-Attn & \cmark & \cmark & \underline{0.902} & \underline{0.943} & \underline{0.089} & \underline{94.00} \\
    + Color Transfer & \cmark & \cmark & \textbf{0.904} & \textbf{0.960} & \textbf{0.080} & \textbf{94.75} \\
    \bottomrule
\end{tabular}
\label{tab:ablation}
\end{table}

\begin{figure}[t]
    \centering
    \includegraphics[width=\linewidth]{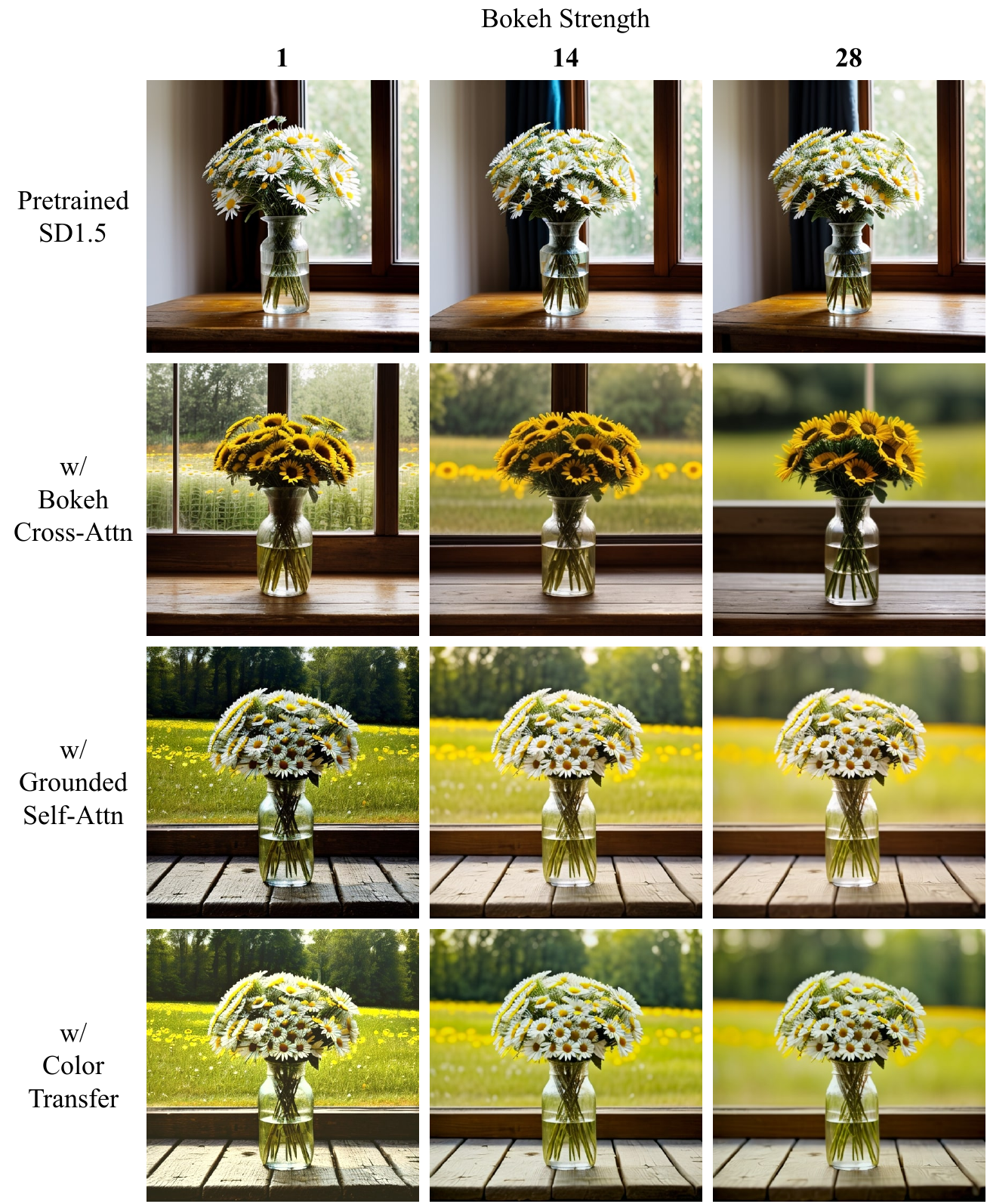}
    \caption{Ablation of the proposed modules of Bokeh Diffusion.}
    \Description{}
    \label{fig:ablation}
\end{figure}

\subsubsection{Training data.}
We train on three different settings: \emph{(i)} only in-the-wild (ITW) data, \emph{(ii)} only synthetic (SYN) blur augmentations, and \emph{(iii)} a dataset combining both. \tabref{ablation} shows the mixed setting achieves the best overall accuracy while also improving consistency compared to training solely on SYN data. Training on ITW data alone results in high consistency but lower accuracy. We hypothesize that the diversity and potential noise in real image annotations make it hard for the model to learn a clear correlation between the conditioning signal and the bokeh effect. As a result, the model learns to apply only a limited bokeh range. Conversely, training exclusively on SYN data allows for fast convergence but degrades the generation quality and generalization ability of the model.

\subsubsection{Imposing scene consistency.}
We ablate key components of our method to understand their contribution to achieving scene-consistent bokeh control. Starting with the bokeh cross-attention module alone, we observe improved defocus control, indicating that explicitly conditioning on a physical blur parameter already enables the model to modulate depth-of-field meaningfully. However, without additional structure guidance, content shifts occur across different blur levels. Adding grounded self-attention significantly enhances scene consistency, ensuring that spatial layout remains aligned as the blur strength varies. This validates our design of using pivot-based guidance to anchor the structure of the generated image. We note that grounded self-attention introduces moderate computational overhead, increasing runtime per sample by $\sim$76\% and peak memory usage by $\sim$33\% (see supplementary material for detailed measurements). Lastly, while the color transfer step is optional, it further improves visual coherence by harmonizing lighting and tone, without interfering with the defocus itself. \figref{ablation} provides a visual example of the contribution of each component.

\begin{figure*}[t]
    \centering
    \centering
    \includegraphics[width=\linewidth]{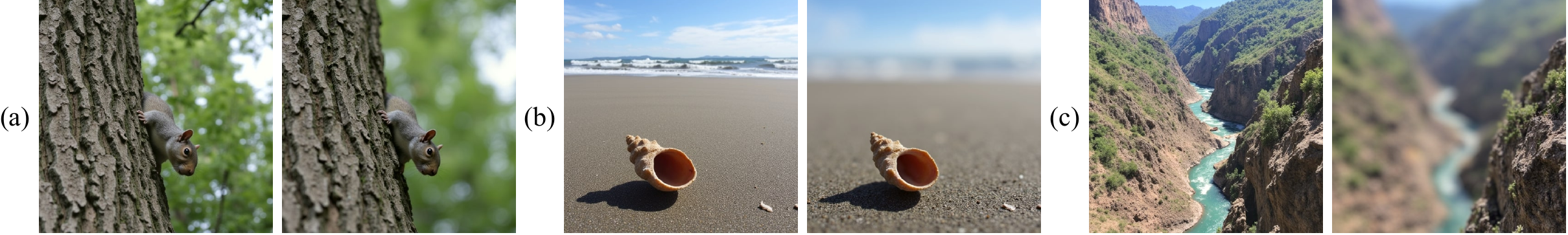}
    \caption{Failure cases of our method: (a) Failure to generate an all-in-focus image (left) in uncommon settings. Long-lens wildlife photos almost always exhibit strong background blur, such sharp examples are under-represented in training, so the model retains residual bokeh; (b) Over-sharpened foreground (right); and (c) When no clear foreground is present (left), the model occasionally hallucinates spurious foreground elements if asked to increase bokeh (right).}
    \Description{}
    \label{fig:limitation}
\end{figure*}

\section{Limitations and Discussion}
\label{sec:limitations}

\figref{limitation} illustrates a few representative failure cases of our method, reflecting data-driven tendencies. Our approach relies on in-the-wild images enriched with inferred parameters (including metric depth and a heuristic focus distance), which inevitably introduces approximation error. Beyond this, some observed failures appear rooted in common photographic conventions (\figref{limitation}a, \figref{limitation}b). Additionally, since most training scenes contain a clear foreground subject, the model may over-assume such structure and hallucinate foreground content when none exists (\figref{limitation}c). These observations motivate promising future extensions. A key direction is enabling flexible focal-plane control, allowing the sharp region to slide through the scene. We also envision conditioning on other photographic settings (\eg, exposure) and extending our method to image-to-image diffusion for improved real-image editing use cases.

\section{Conclusion}
\label{sec:conclusion}

In this work, we introduced Bokeh Diffusion, a text-to-image framework that enables precise defocus blur control while maintaining scene consistency. We integrate in-the-wild images with synthetic blur augmentations to balance realism and robustness, while our grounded self-attention mechanism enforces structural consistency across blur levels. Experiments show our method surpasses baselines in both accuracy and scene consistency, offering flexible, lens-like bokeh adjustments. We hope this work inspires further research into fine-grained photographic controls in generative models.

\begin{acks}
This research is supported by NTU SUG-NAP and is also supported by cash and in-kind funding from NTU S-Lab and industry partner(s).
\end{acks}

\bibliographystyle{ACM-Reference-Format}
\bibliography{main}

\clearpage

\clearpage

\noindent{\fontsize{17.28}{20}\selectfont{Supplementary Material}\par}

\appendix

\section{Additional Method Details}

\subsection{Dataset}

\subsubsection{Foreground reference and focus distance}

Since EXIF data rarely includes an explicit focus distance $S_1$, we approximate it by taking the median depth of the nearest portion of the foreground region. When the foreground depth range is small, we treat the entire foreground at a single $S_1$ to ensure a sharp subject.

\subsubsection{Dataset alignment}

\begin{figure}[h]
    \centering
    \includegraphics[width=\linewidth]{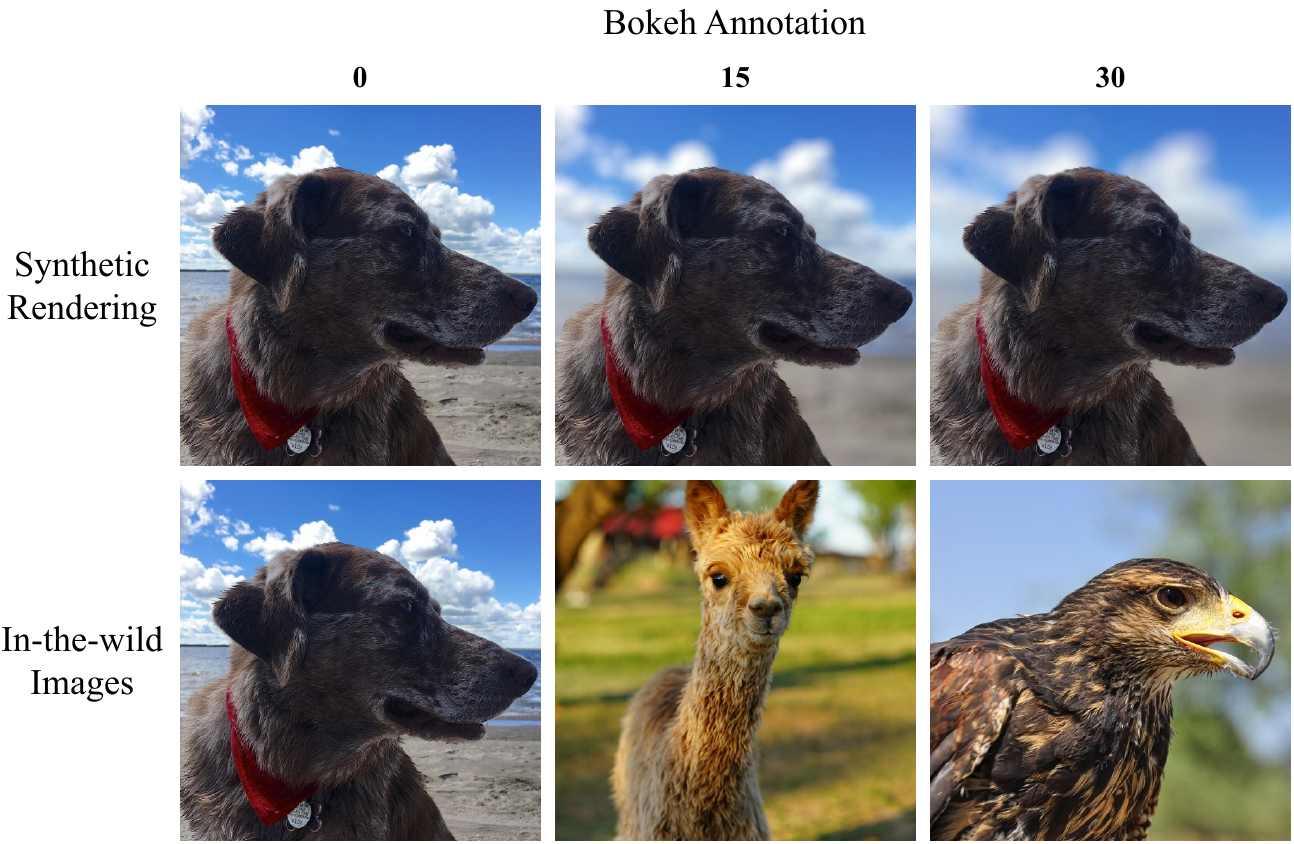}
    \caption{Visualization of bokeh annotation alignment between collected in-the-wild images and synthetic bokeh augmentations using BokehMe~\cite{pengBokehMeWhenNeural2022} at 512$\times$512 resolution. Dog \textcopyright{} Alan Levine (Flickr), Llama \textcopyright{} Alan Kyker (Flickr), Hawk \textcopyright{} Frayle (Flickr).}
    \Description{}
    \label{fig:datasetvs}
\end{figure}

Given the unified bokeh annotation $K$ scale, \figref{datasetvs} illustrates the alignment between the synthetic blur augmentations and the in-the-wild defocus patterns. The figure visually compares examples from both data sources, demonstrating that the synthetic approach closely replicates the depth-of-field characteristics observed in real photographs. This alignment validates our method for unifying the blur parameter $K$ across diverse data, ensuring that the model learns consistent depth-of-field control.

\subsubsection{Subject diversity}

\begin{figure}[ht]
    \centering
    \includegraphics[width=\linewidth]{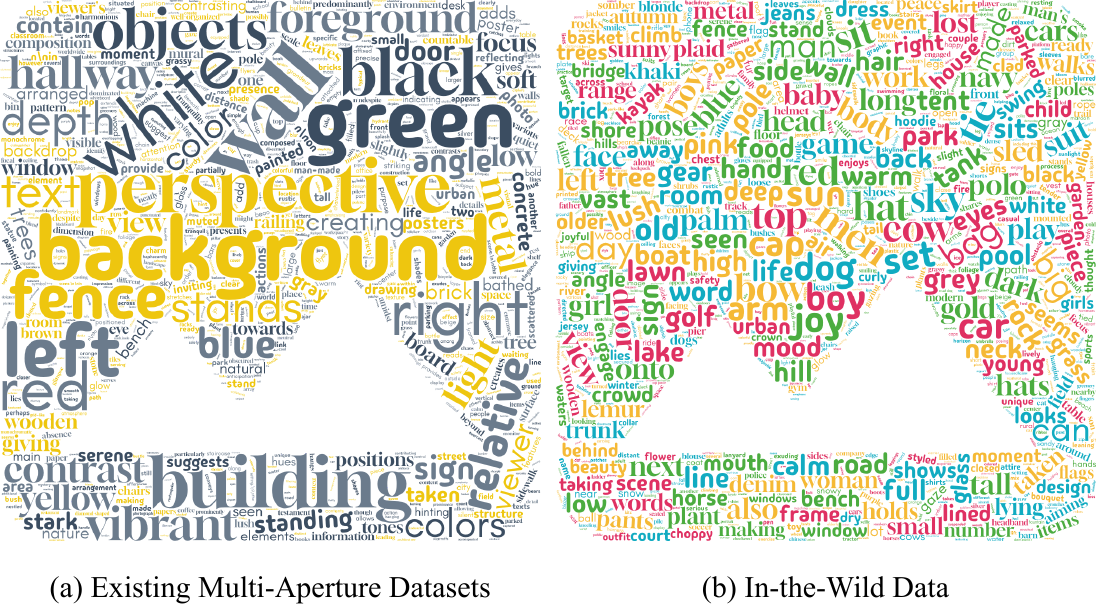}
    \caption{Wordclouds of captions extracted using a vision-language model for (a) DPDD~\cite{abuolaim2020defocus} and (b) our in-the-wild dataset. Our dataset exhibits greater subject diversity, while the DPDD dataset focuses on static scenes with inanimate objects.}
    \Description{}
    \label{fig:wordcloud}
\end{figure}

To further support our claim that existing multi-aperture datasets are limited in content diversity, \figref{wordcloud} visualizes the distribution of scene content using wordclouds generated from vision-language model (VLM) captions. Specifically, we compare the captions extracted from the DPDD dataset~\cite{abuolaim2020defocus}---a prominent multi-aperture dataset---with those from our collected in-the-wild images. As seen in the visualization, DPDD predominantly features static, inanimate scenes such as indoor objects and tabletop arrangements. In contrast, our in-the-wild dataset spans a much broader range of subjects, including dynamic environments, natural scenes, and human-centered contexts. This diversity is crucial for training a generative model capable of realistic and generalizable bokeh control across varied prompts.

\subsection{Localized cross-attention.}

In addition to the decoupled cross-attention mechanism for bokeh condition injection, we adapt a localization strategy from \citet{xiaoFastComposerTuningFreeMultiSubject2023} to steer the bokeh cross-attention toward the background region during training. Specifically, we downsample each foreground mask and measure the average bokeh cross-attention in the foreground and background. Analogously to \citet{xiaoFastComposerTuningFreeMultiSubject2023}, we then add a small regularization term that encourages higher attention values in the background and lower in the foreground. While this localization loss does not fully resolve scene inconsistency, it acts as a soft constraint that nudges the model toward learning more interpretable and spatially aware defocus behavior, providing mild supervision that complements the rest of the framework.

\subsection{Algorithm}

The pseudo-code for scene-consistent bokeh-conditioned sampling in the SD1.5~\cite{rombachHighResolutionImageSynthesis2022} version of our method (UNet-based) is provided in Alg.~\ref{alg:sampling}. The FLUX~\cite{labsFLUX2024} version of our method is equivalent, aside from the fact that the decoupled bokeh cross-attention is perfomed in all layers and the number of default sampling and grounding steps is 30 and 18, respectively.

\begin{algorithm}[t]
\caption{\textsc{Scene-Consistent Bokeh Sampling (SD1.5)}}
\label{alg:sampling}
\DontPrintSemicolon     
\KwIn{Prompt $\mathcal{P}$; pivot bokeh $\mathcal{K}_{\mathrm{piv}}$; target bokeh $\mathcal{K}_{\mathrm{tgt}}$; our finetuned T2I diffusion model $\varepsilon_{\theta}$ with $L$ transformer blocks; decoder $\mathcal{D}$; total sampling steps $T\,(\text{default}\,50)$; grounding cutoff $G\,(\text{default}\,30)$.}
\KwOut{Image $x^{\mathrm{tgt}}$ matching $\mathcal{K}_{\mathrm{tgt}}$ while preserving the pivot scene of $x^{\mathrm{piv}}$.}

\textbf{Initialise} $z_T^{\mathrm{piv}}\!\sim\!\mathcal{N}(0,1)$, $z_T^{\mathrm{tgt}}\leftarrow z_T^{\mathrm{piv}}$.

\For{$t=T$ \KwTo $T-G+1$}{
  \For{$\ell=1$ \KwTo $L$}{
    $Q^{\ell}_{\mathrm{piv}},K^{\ell}_{\mathrm{piv}},V^{\ell}_{\mathrm{piv}}
        \leftarrow\varepsilon_{\theta}\bigl(z_t^{\mathrm{piv}},t,
        \mathcal{P},\mathcal{K}_{\mathrm{piv}}\bigr)$\;
    $Q^{\ell}_{\mathrm{tgt}},K^{\ell}_{\mathrm{tgt}},V^{\ell}_{\mathrm{tgt}}
        \leftarrow\varepsilon_{\theta}\bigl(z_t^{\mathrm{tgt}},t,
        \mathcal{P},\mathcal{K}_{\mathrm{tgt}}\bigr)$\;

    $A^{\ell}\leftarrow
      \text{Attention}\bigl(Q^{\ell}_{\mathrm{piv}},
      [K^{\ell}_{\mathrm{tgt}},K^{\ell}_{\mathrm{piv}}],
      [V^{\ell}_{\mathrm{tgt}},V^{\ell}_{\mathrm{tgt}}]\bigr)$\;

    \If{$\ell\in\mathcal{L}_{\mathrm{deep}}$}{   
      $h^{\ell}_{\mathrm{piv}}\leftarrow h^{\ell}_{\mathrm{piv}}+\lambda\,\text{Attention}\bigl(Q^{\ell}_{\mathrm{piv}},K^{\ell}_{b-\mathrm{piv}},V^{\ell}_{b-\mathrm{piv}}\bigr)$\;
      
      $h^{\ell}_{\mathrm{tgt}}\leftarrow h^{\ell}_{\mathrm{tgt}}+\lambda\,\text{Attention}\bigl(Q^{\ell}_{\mathrm{tgt}},K^{\ell}_{b-\mathrm{tgt}},V^{\ell}_{b-\mathrm{tgt}}\bigr)$\;
    }
  }
  $z_{t-1}^{\mathrm{piv}}\leftarrow\varepsilon_{\theta}\bigl(z_t^{\mathrm{piv}},t,\mathcal{P},\mathcal{K}_{\mathrm{piv}}\bigr)$\;
  
  $z_{t-1}^{\mathrm{tgt}}\leftarrow\varepsilon_{\theta}\bigl(z_t^{\mathrm{tgt}},t,\mathcal{P},\mathcal{K}_{\mathrm{tgt}},A\bigr)$\;
}

\For{$t=T-G$ \KwTo $1$}{
  $z_{t-1}^{\mathrm{piv}}\leftarrow\varepsilon_{\theta}\bigl(z_t^{\mathrm{piv}},t,\mathcal{P},\mathcal{K}_{\mathrm{piv}}\bigr)$\;
      
  $z_{t-1}^{\mathrm{tgt}}\leftarrow \varepsilon_{\theta}\bigl(z_t^{\mathrm{tgt}},t,\mathcal{P},\mathcal{K}_{\mathrm{tgt}}\bigr)$\;
}

$x^{\mathrm{piv}}\leftarrow\mathcal{D}\bigl(z_0^{\mathrm{piv}}\bigr)$

$x^{\mathrm{tgt}}\leftarrow\mathcal{D}\bigl(z_0^{\mathrm{tgt}}\bigr)$\;

\Return $x^{\mathrm{tgt}}$\;
\end{algorithm}

\section{Additional Experimental Details}

\subsection{Implementation Details}

During training, each step samples either a real batch or a synthetic batch with equal probability (50\%). In-the-wild batches use the conventional self-attention. In contrastive batches, we employ our grounded self-attention mechanism, where a single image in the batch is randomly selected as pivot for the others. Training data captions are taken from HuggingFace\footnote{\href{https://huggingface.co/datasets/CaptionEmporium/flickr-megalith-10m-internvl2-multi-caption}{CaptionEmporium/flickr-megalith-10m-internvl2-multi-caption}}. For the SD1.5~\cite{rombachHighResolutionImageSynthesis2022} version of our model, we use the AdamW optimizer~\cite{loshchilovDecoupledWeightDecay2018} with learning rate $1e^{-4}$ and total batch size 80. Following \citet{guttenbergDiffusionOffsetNoise2023}, we apply a noise offset of $0.04$. We also set a min-SNR of $2.0$~\cite{hangEfficientDiffusionTraining2023}.
We train for around 12K steps, taking approximately 3 hours on 4 NVIDIA RTX 4090 GPUs.
At inference, we usually run 50 denoising steps with DPM-Solver~\cite{luDPMSolverFastODE2022a}; of these, about 30 steps employ grounded self-attention. Inference results for this version are based on community checkpoints\footnote{\href{https://huggingface.co/SG161222/Realistic\_Vision\_V5.1\_noVAE}{SG161222/Realistic\_Vision\_V5.1}}\footnote{\href{https://huggingface.co/cyberdelia/CyberRealistic}{cyberdelia/CyberRealistic}}. For the FLUX~\cite{labsFLUX2024} version, we train a LoRA~\cite{hu2022lora} of rank 128 using the Prodigy optimizer~\cite{mishchenkoProdigyExpeditiouslyAdaptive2024} with learning rate $1$ and total batch size 48. We also train this model for around 12K steps. At inference, we run 30 denoising steps with Euler~\cite{karrasElucidatingDesignSpace2022a}; of these about 18 steps employ grounded self-attention. Both models are trained at 512$\times$512 resolution.

\subsection{Baseline Details}

To introduce bokeh conditions in the pretrained T2I baselines, we append short descriptors to the text prompt. For the finetuned baselines, CSaT and GenPhoto, we follow their respective conditioning protocols, with some adaptations for fair comparison. CSaT~\cite{fangCameraSettingsTokens2024} conditions generation on explicit camera intrinsics, such as aperture, focal length, and ISO. In our evaluation, we fix all parameters except aperture and vary only the F-number to control the depth-of-field. According to the thin-lens equation (see main paper), the blur radius per unit disparity is inversely proportional to the f-number. Therefore, we apply bokeh transitions in CSaT by adjusting aperture values inversely: for example, doubling the target bokeh strength corresponds to halving the F-number. GenPhoto~\cite{yuanGenerativePhotographySceneConsistent2024} shares the same defocus conditioning parameter as our method (blur radius in pixels per unit disparity). However, it is trained at a different image resolution. Since this parameter is defined in pixel space, we proportionally scale the bokeh values to ensure consistent visual blur across resolutions.

\subsection{GPT-4o Evaluation}

We leverage GPT-4o~\cite{openaiGPT4oSystemCard2024} as a perceptual evaluator for two complementary metrics: \emph{bokeh accuracy} and \emph{scene consistency}. All evaluations are conducted via OpenAI's vision-language API, where images are passed alongside structured natural language instructions, and the model returns scalar outputs. We use a batch-based protocol, with images presented in context.

\subsubsection{Bokeh accuracy}

To assess how well the bokeh level in each image aligns with the intended defocus strength, we prompt GPT-4o to estimate a bokeh strength score for each image in a batch, within the given defocus blur range. The instruction used is as follows:
\begin{quote}
\emph{We have several images below, each with a different bokeh strength. For each image, please provide a numeric bokeh strength score between 0 and 30. A very low score means the image is in deep focus, while a high score means very strong defocus blur in the background. You should return only the scores and nothing else, in a single line, comma-separated, in the same order as the images.}
\end{quote}
The returned scores are then compared with the respective conditioning sequence using Pearson correlation to compute our GPT-based bokeh alignment metric.

\subsubsection{Scene consistency}

To evaluate how well a method preserves the underlying scene structure across varying bokeh levels, we present a batch of images and ask GPT-4o to provide a single scene consistency score. The instruction is:
\begin{quote}
\emph{We have several images below of presumably the same scene but with varying depths-of-field. Please provide a single numeric score from 0 to 100 indicating how consistent the scene/subjects are across all these images, ignoring pure defocus blur differences but penalizing any other changes. A score of 100 means all images show precisely the same structure, while 0 means they're drastically changed in arrangement. Any minor variations in any of the images aside from bokeh should also be penalized. The only numeric score you should return is the scene consistency score.}
\end{quote}
This score serves as an additional perceptual consistency signal, complementing traditional metrics such as LPIPS.

\subsection{User Study}

\begin{figure}[t]
    \centering
    \includegraphics[width=\linewidth]{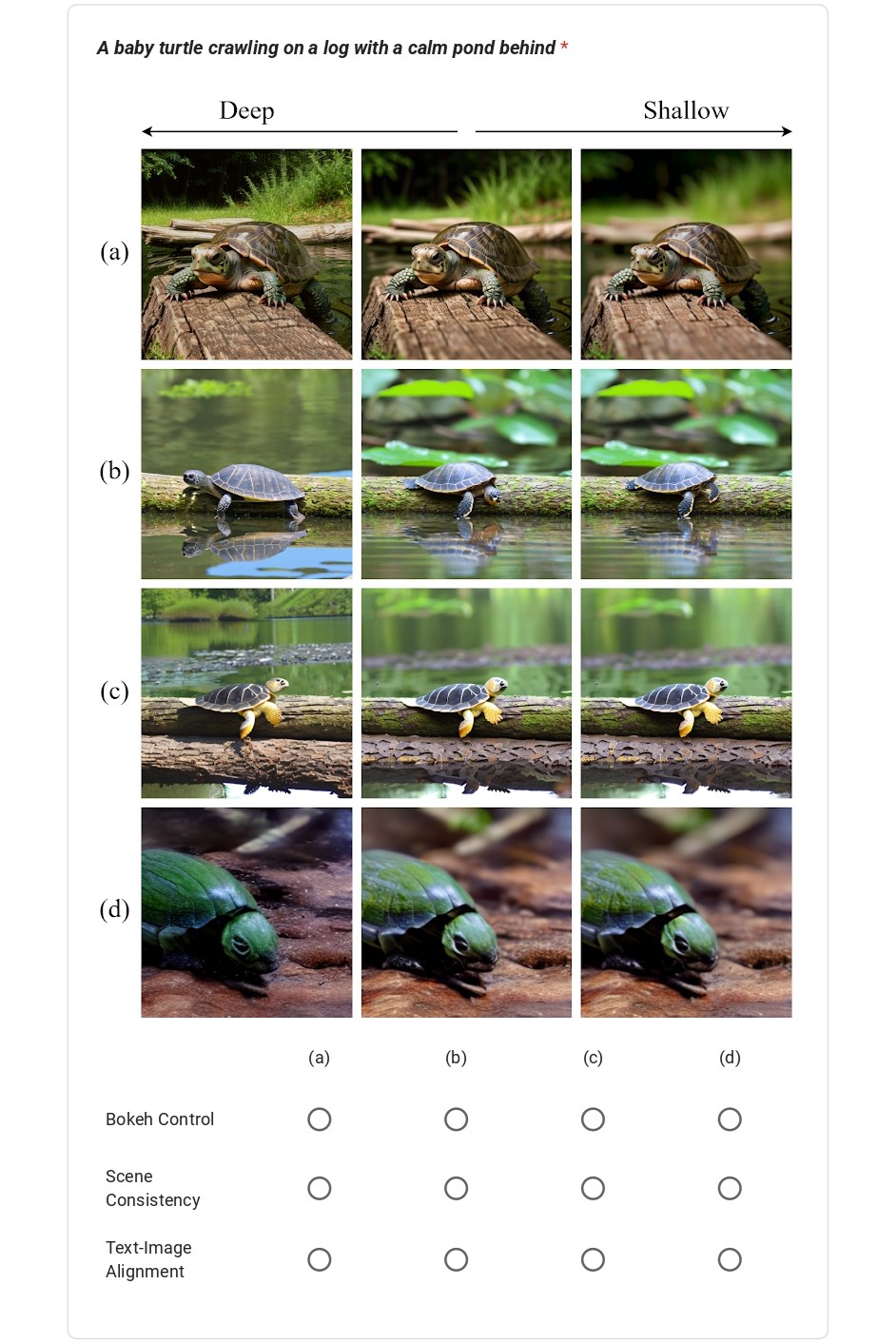}
    \caption{Example user study question. Participants were shown a text prompt along with image sequences from four anonymized methods (randomized order), and asked to choose the best sequence along three evaluation criteria.}
    \Description{}
    \label{fig:user_example}
\end{figure}

We conducted a user study with 20 participants to evaluate bokeh control, scene consistency, and text-image alignment. Participants were presented with 15 prompts accompanied by outputs from four anonymized methods. The outputs were arranged in grids of image sequences (from deep to shallow depth-of-field, left to right). For each scene, the order of methods was randomized for anonymity. The following instruction was shown to users at the start:
\begin{quote}
\emph{Each question will start with a text prompt and include sequences of images from 4 different methods (random order). Each image sequence represents the same scene going from a deep to a shallow depth-of-field (left to right). You should choose the best result across 3 dimensions: \textbf{Bokeh Control:} Choose the option with the most accurate depth-of-field sequence: a deep depth-of-field should make contents appear in-focus and sharp (left); a shallow depth-of-field should create more background blur (right). \textbf{Scene Consistency:} Choose the option with the most consistent scene across different depth-of-field conditions (i.e., the image contents are stable from left to right). \textbf{Text-Image Alignment:} Choose the option that best follows the text prompt.}
\end{quote}
An example interface shown to participants is provided in \figref{user_example}.

\section{Additional Ablations}

\subsection{Image Color Transfer}

\begin{figure}[t]
    \centering
    \includegraphics[width=\linewidth]{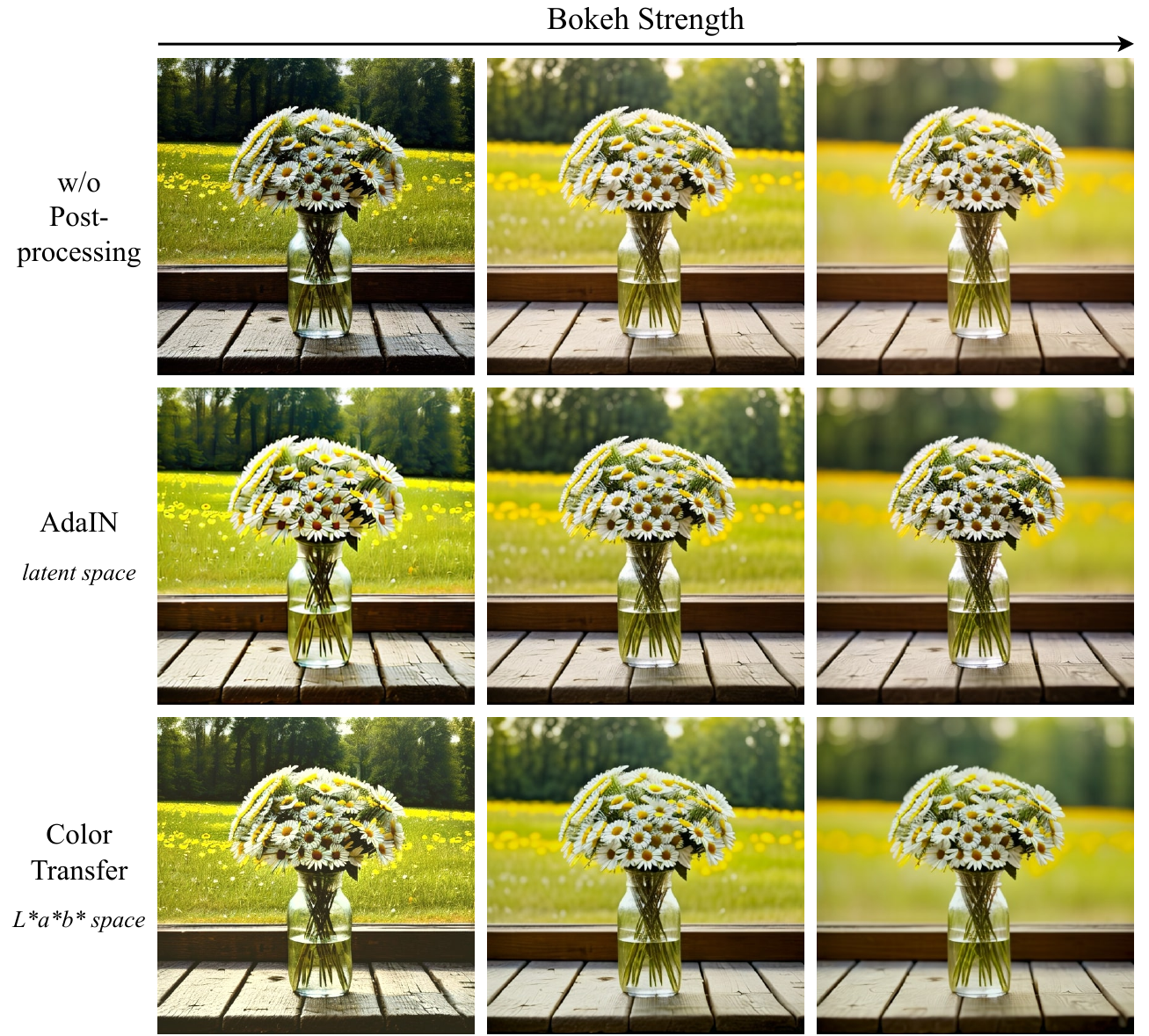}
    \caption{Comparison of post-processing methods to mitigate color and lighting biases introduced by bokeh variation. Best viewed zoomed in.}
    \Description{}
    \label{fig:post}
\end{figure}

As described in the main paper, a post-processing step is optionally applied to harmonize color and lighting across defocus levels. This helps mitigate subtle biases that can emerge when blur strength changes, especially under varying exposure in the training data. This phenomenon has also been reported in related diffusion model applications~\cite{yangRerenderVideoZeroShot2023}, where it is commonly addressed using AdaIN~\cite{huangArbitraryStyleTransfer2017} in the latent space. Instead, inspired by the classic color-transfer method of~\citet{reinhardColorTransferImages2001}, we operate in the $L^*a^*b^*$ color space. Figure~\ref{fig:post} compares three versions of a target bokeh image: (1) without color transfer, (2) using AdaIN~\cite{huangArbitraryStyleTransfer2017} in the latent space, and (3) using the method based on classic $L^*a^*b^*$ statistics matching~\cite{reinhardColorTransferImages2001}. While AdaIN effectively matches color and contrast to some extent, it also alters local variance (better noticeable in left column). In contrast, the $L^*a^*b^*$-based method matches only the per-channel mean and variance in color space, leaving local detail structure and defocus effects intact. As a result, it preserves the intended depth-of-field while reducing color inconsistencies across scenes.

\subsection{Grounding Steps}

\begin{figure*}[p]
    \centering
    \includegraphics[width=\textwidth]{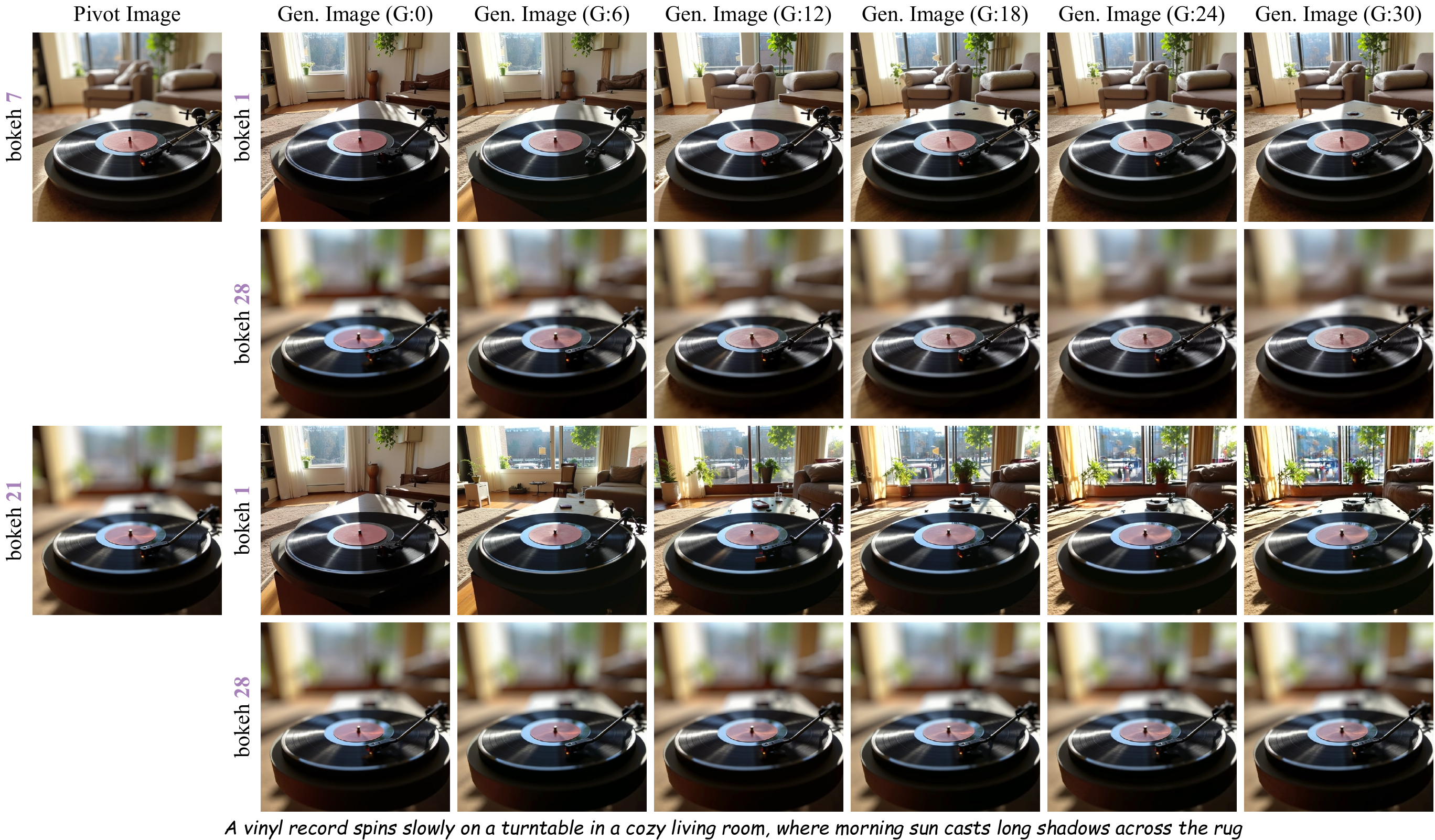} \\
    \vspace{-3mm}
    \caption{Effect of varying number of grounding steps during sampling in our FLUX-based model (out of 30 total).}
    \Description{}
    \label{fig:additional}
\end{figure*}

To study the impact of the number of grounding steps used during sampling, we perform an ablation on our FLUX-based model, which uses 30 total denoising steps. As shown in \figref{additional}, the model is robust to a range of grounding configurations. Our default setting uses 18 grounding steps (60\% of the sampling process), which mirrors the configuration used in the SD1.5 version and provides a good trade-off between scene consistency and flexible defocus control. Using too few grounding steps may lead to structural drift or inconsistencies in subject contours. Conversely, applying grounding for too many steps can overly constrain the model, occasionally resulting in artifacts when reducing blur or suppressing the achievable bokeh range when increasing it.

\subsection{Runtime and Memory Impact of Grounded Self-Attention}
\label{sec:runtime}

We benchmarked runtime and GPU memory consumption on an RTX~4090 GPU in \texttt{fp16}. Table~\ref{tab:gsa_runtime} reports the per-sample runtime and peak memory usage for the SD1.5 baseline and SD1.5 augmented with grounded self-attention (GSA).
\setlength{\tabcolsep}{5pt}
\begin{table}[t]
\caption{Runtime and memory usage of grounded self-attention (GSA).}
\centering
\begin{tabular}{l|cc}
    \toprule
    Model & Runtime/sample (s) & VRAM (GB) \\ \midrule
    SD1.5 & 1.248 & 3.772 \\
    SD1.5 + GSA & 2.193 & 5.032 \\
    \bottomrule
\end{tabular}
\label{tab:gsa_runtime}
\end{table}

\section{Additional Visual Results}

\begin{figure*}[p]
    \centering
    \includegraphics[width=\textwidth]{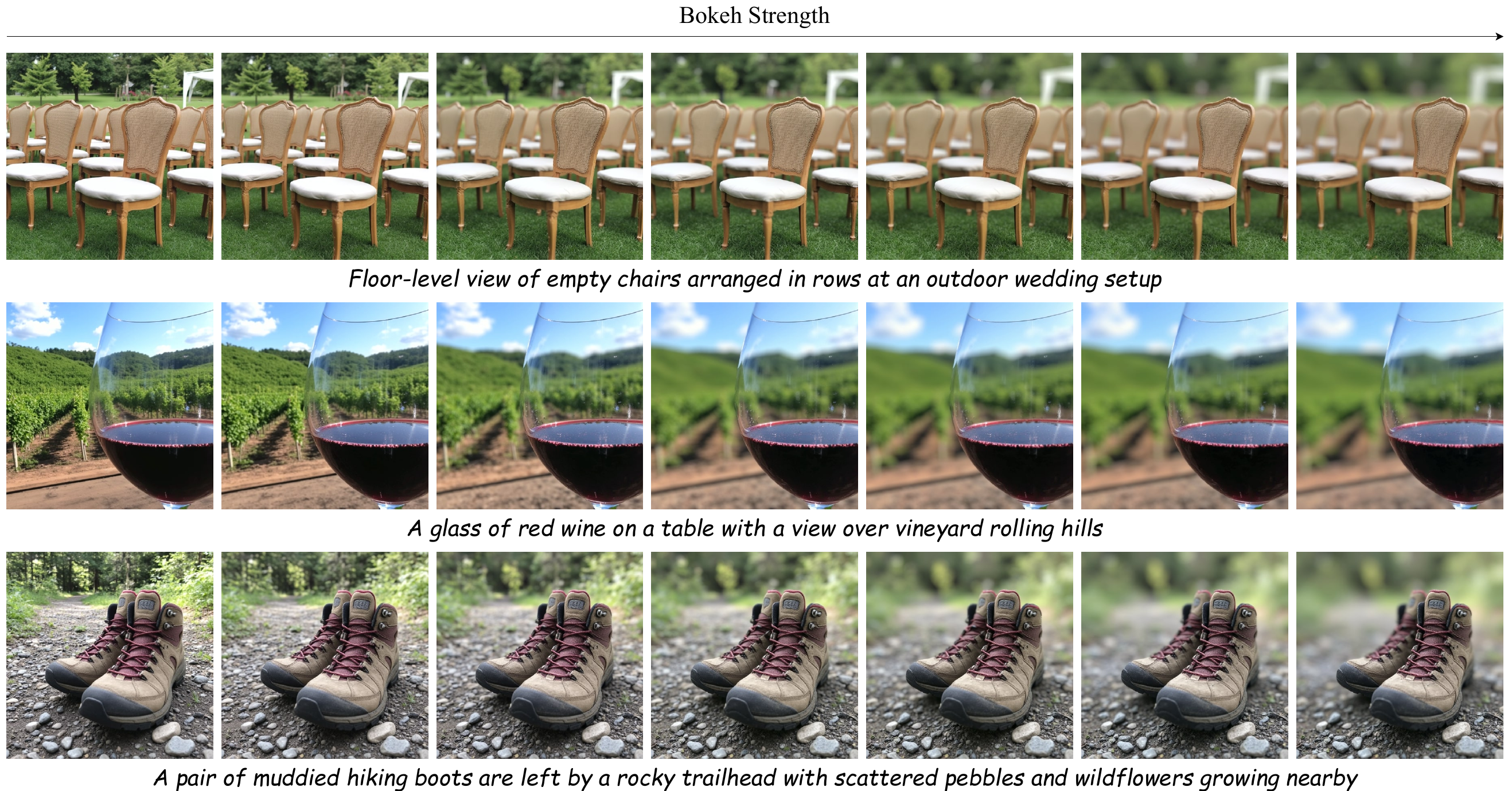} \\
    \vspace{-2mm}
    \caption{Continuous bokeh variation across fixed scenes. Our method allows smooth and physically grounded transitions between deep focus and shallow focus, producing consistent visual structure across intermediate blur levels.}
    \Description{}
    \label{fig:continuous}
\end{figure*}

\begin{figure*}[p]
    \centering
    \includegraphics[width=0.98\textwidth]{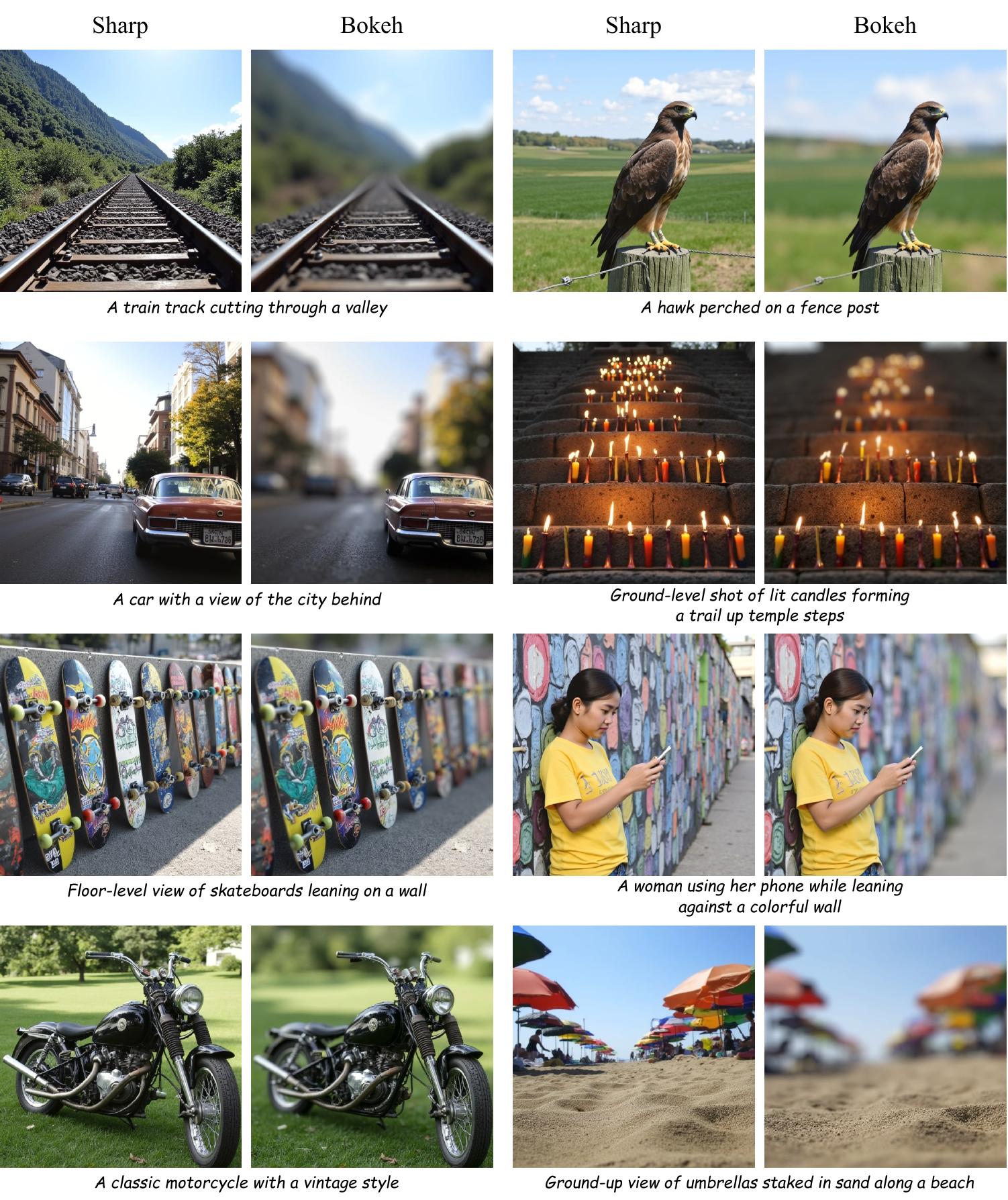} \\ 
    \vspace{-2mm}
    \caption{Additional visual results for scene-consistent bokeh-conditioned generation from Bokeh Diffusion on FLUX~\cite{labsFLUX2024}.}
    \Description{}
    \label{fig:additional-flux}
\end{figure*}

\begin{figure*}[p]
    \centering
    \includegraphics[width=0.98\textwidth]{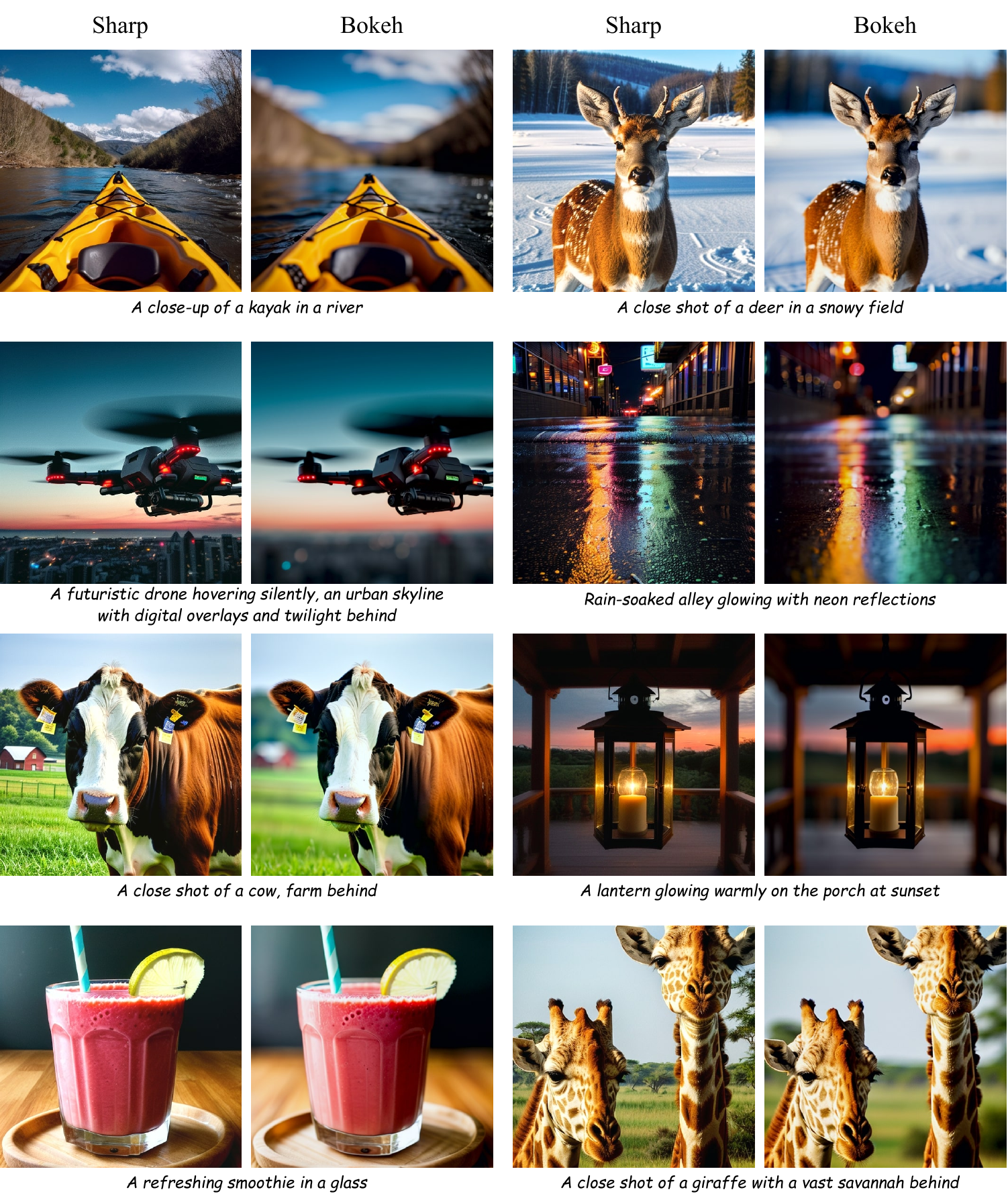} \\ 
    \vspace{-2mm}
    \caption{Additional visual results for scene-consistent bokeh-conditioned generation from Bokeh Diffusion on SD1.5~\cite{rombachHighResolutionImageSynthesis2022}.}
    \Description{}
    \label{fig:additional-sd}
\end{figure*}

\begin{figure*}[p]
    \centering
    \includegraphics[width=0.96\textwidth]{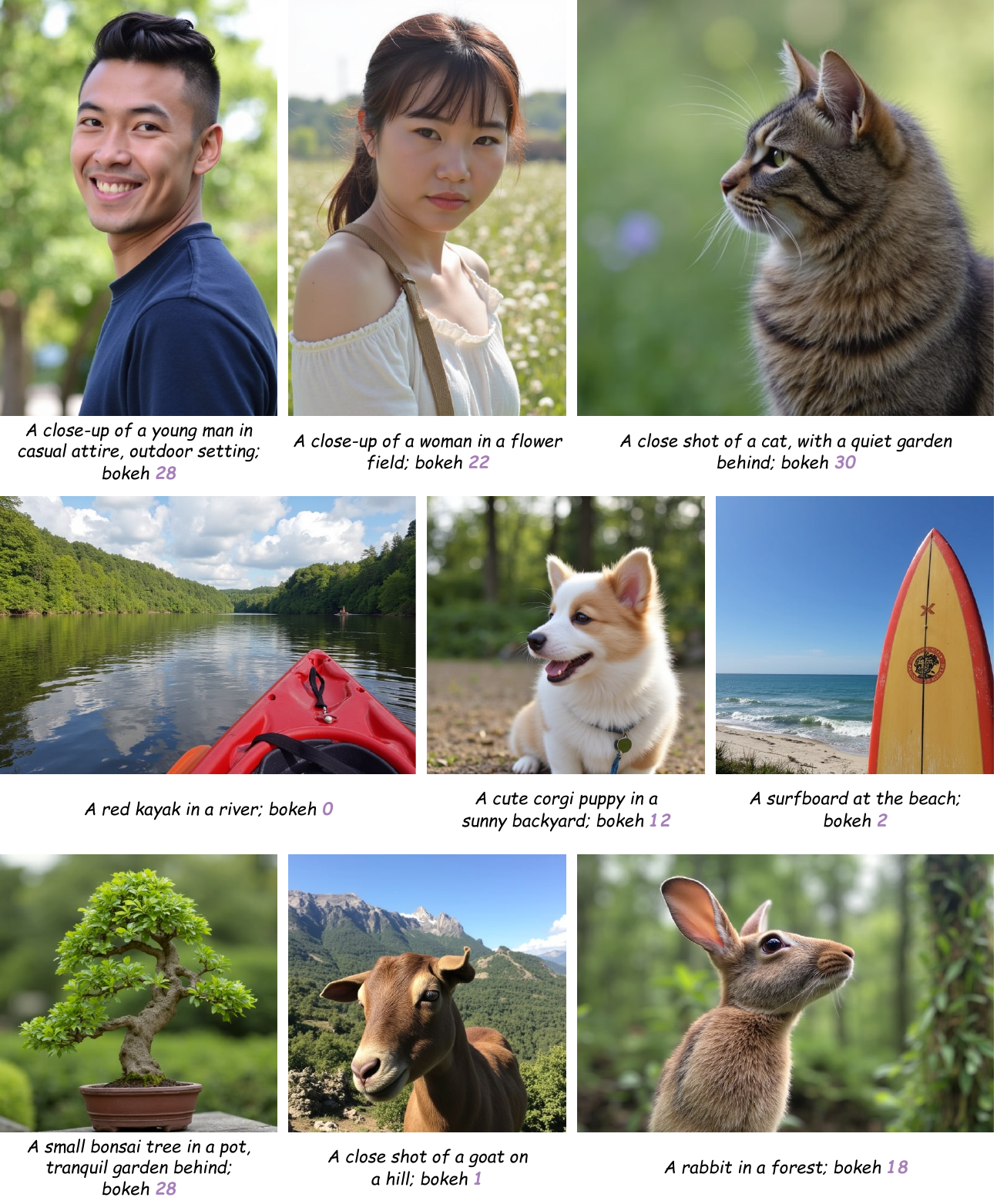} \\ 
    \vspace{-3mm}
    \caption{Additional visual results for unbounded bokeh-conditioned generation from Bokeh Diffusion on FLUX~\cite{labsFLUX2024}.}
    \Description{}
    \label{fig:additional-unbounded-flux}
\end{figure*}

\begin{figure*}[p]
    \centering
    \includegraphics[width=0.96\textwidth]{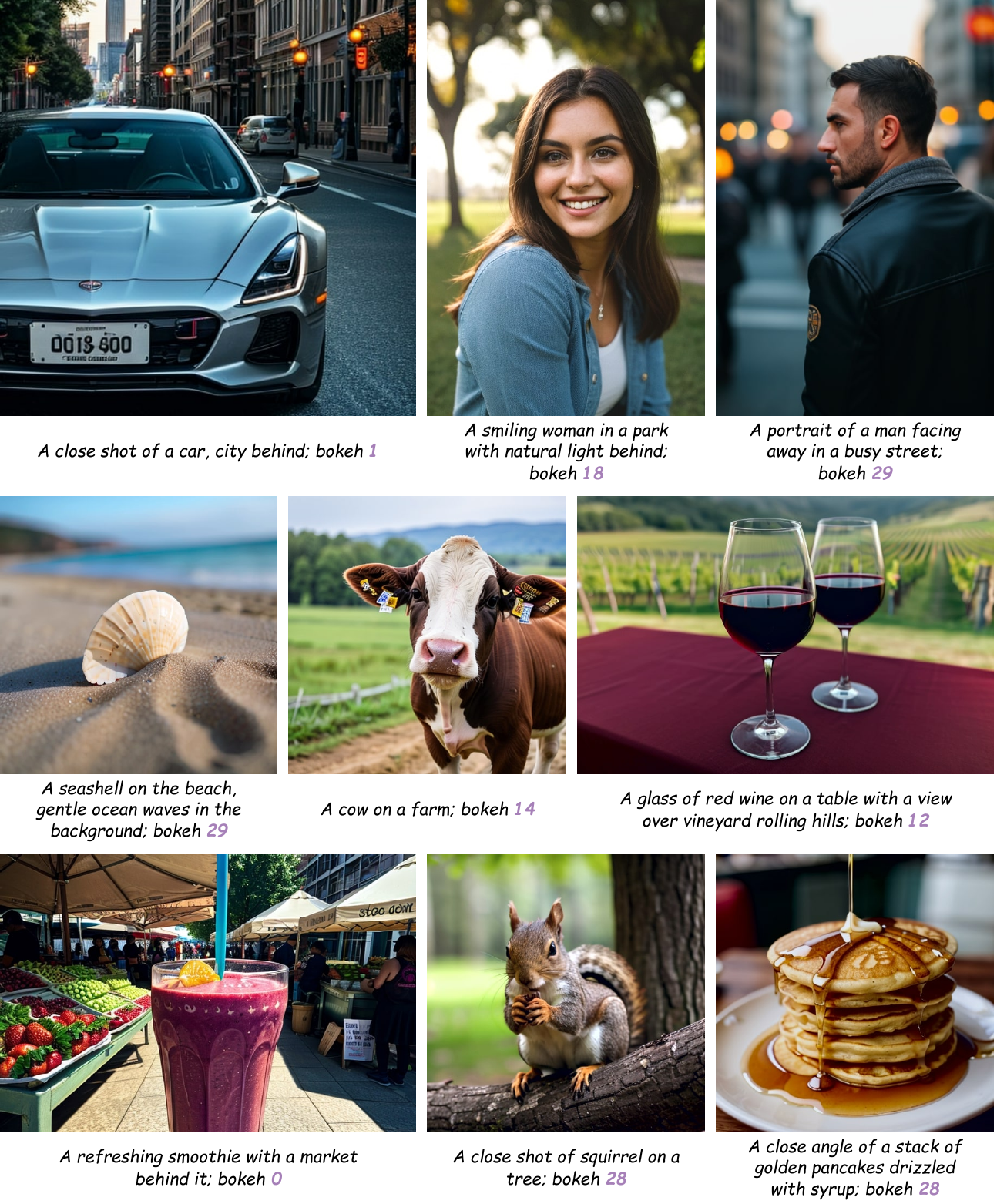} \\ 
    \vspace{-3mm}
    \caption{Additional visual results for unbounded bokeh-conditioned generation from Bokeh Diffusion on SD1.5~\cite{rombachHighResolutionImageSynthesis2022}.}
    \Description{}
    \label{fig:additional-unbounded-sd}
\end{figure*}

We provide additional qualitative results to highlight the versatility and consistency of Bokeh Diffusion under varying blur conditions. \figref{continuous} illustrates a continuous progression of bokeh strength for a fixed scene, emphasizing the smooth, physically plausible control enabled by our method. \figref{additional-flux} and \figref{additional-sd} present more examples generated with our FLUX- and SD1.5-based models, respectively. We also show diverse unbounded bokeh control examples in \figref{additional-unbounded-flux} and \figref{additional-unbounded-sd}. For dynamic visualizations, we encourage readers to view the additional supplementary material, which includes video animations demonstrating continuous bokeh transitions across multiple scenes.

\end{document}